\documentclass[useAMS,usenatbib]{mn2e}
\usepackage{epsfig}
\usepackage{amssymb,amsmath}
\usepackage{subfigure}

\title[BLASTPol HWP]{Empirical modelling of the BLASTPol achromatic half-wave plate for precision submillimetre polarimetry}

\author[Moncelsi, et al.] {Lorenzo Moncelsi$^{1,2}$\thanks{E-mail:
moncelsi@caltech.edu}, Peter A.~R.~Ade$^{2}$, Francesco E.~Angil\`{e}$^{3}$, Steven J.~Benton$^{4}$, \newauthor Mark J.~Devlin$^{3}$, Laura M.~Fissel$^{5}$, Natalie N.~Gandilo$^{5}$, Joshua O.~Gundersen$^{6}$, \newauthor Tristan G.~Matthews$^{7}$, C.~Barth Netterfield$^{4,5}$, Giles Novak$^{7}$, David Nutter$^{2}$, Enzo \newauthor Pascale$^{2}$, Fr\'{e}d\'{e}rick Poidevin$^{8}$, Giorgio Savini$^{8}$, Douglas Scott$^{9}$, Juan Diego~Soler$^{5}$, \newauthor Locke D.~Spencer$^{2,10}$, Matthew D.~P.~Truch$^{3}$, Gregory S.~Tucker$^{11}$, Jin Zhang$^{2}$\\
$^{1}$ California Institute of Technology, 1200 E. California Blvd., Pasadena, CA\ 91125, USA \\
$^{2}$ Department of Physics and Astronomy, Cardiff University, 5 The Parade, Cardiff, CF24~3AA, UK \\
$^{3}$ Department of Physics and Astronomy, University of Pennsylvania, 209 South 33rd Street, Philadelphia, PA\ 19104, USA\\
$^{4}$ Department of Physics, University of Toronto, 60 St. George Street, Toronto, ON\ M5S~1A7, Canada \\
$^{5}$ Department of Astronomy and Astrophysics, University of Toronto, 50 St. George Street, Toronto, ON\ M5S~3H4, Canada \\
$^{6}$ Department of Physics, University of Miami, 1320 Campo Sano Drive, Coral Gables, FL\ 33146, USA \\
$^{7}$ Department of Physics and Astronomy, Northwestern University, Evanston, IL\ 60208, USA \\
$^{8}$ Department of Physics and Astronomy, University College London, Gower Street, London WC1E 6BT, UK \\
$^{9}$ Department of Physics and Astronomy, University of British Columbia, 6224 Agricultural Road, Vancouver, BC V6T~1Z1, Canada \\
$^{10}$Department of Physics and Astronomy, University of Lethbridge, 4401
University Drive, Lethbridge, Alberta, Canada, T1K 3M4\\
$^{11}$Department of Physics, Brown University, 182 Hope Street,
Providence, RI\ 02912, USA}

\begin{document}

\date{Accepted . Received ; in original form }

\pagerange{\pageref{firstpage}--\pageref{lastpage}} \pubyear{2012}

\maketitle

\label{firstpage}

\begin{abstract}

A cryogenic achromatic half-wave plate (HWP) for submillimetre
astronomical polarimetry has been designed, manufactured, tested,
and deployed in the Balloon-borne Large-Aperture Submillimeter
Telescope for Polarimetry (BLASTPol). The design is based on the
five-slab Pancharatnam recipe and it works in the wavelength range
200--600\,$\mu$m, making it the broadest-band HWP built to date
at (sub)millimetre wavelengths. The frequency behaviour of the HWP
has been fully characterised at room and cryogenic temperatures
with incoherent radiation from a polarising Fourier transform
spectrometer. We develop a novel empirical model, complementary to
the physical and analytical ones available in the literature, that
allows us to recover the HWP Mueller matrix and phase shift as a
function of frequency and extrapolated to 4\,K. We show that most
of the HWP non-idealities can be modelled by quantifying one
wavelength-dependent parameter, the position of the HWP equivalent
axes, which is then readily implemented in a map-making algorithm.
We derive this parameter for a range of spectral signatures of
input astronomical sources relevant to BLASTPol, and provide a
benchmark example of how our method can yield improved accuracy on measurements of the polarisation angle on the sky at
submillimetre wavelengths.
\end{abstract}


\section{Introduction}

Galactic magnetic fields are believed to play a
crucial role in the evolution of star-forming molecular clouds, perhaps
controlling the rate at which stars are born and even determining their
mass \citep[][]{Crutcher2004b,McKee2007}. 
However, magnetic fields are very difficult to probe on the spatial scales relevant to the star-forming processes, especially within obscured molecular clouds \citep[e.g.,][]{Crutcher2004a,Whittet2008}, hence their influence on star formation has not yet been clearly established observationally.

Zeeman splitting of molecular lines, which allows a direct measurement of the strength of the line-of-sight component of the local magnetic field, has been carried out successfully for a number of molecular cloud cores \citep[e.g.,][]{Crutcher1999a}, though the technique is difficult and often limited to very bright regions \citep{Crutcher2012}. A promising alternative method is to observe clouds with a far-infrared/submillimetre
(FIR/submm) polarimeter
\citep[e.g.,][]{Hildebrand1984,Hildebrand2000,WardThompson2000}. By tracing the
linearly polarised thermal emission from aspherical dust grains aligned with
respect to the local magnetic fields, we can estimate the
direction of the plane-of-the-sky component of the field within
the cloud \citep[][]{Davis1951,Dolginov1976,Lazarian2007}, and its strength via
the Chandrasekhar \& Fermi \citeyearpar[CF;][]{Chandrasekhar1953}
technique, provided that ancillary measurements of the turbulent
motion velocity are available. The observed morphology in submm polarisation maps can also be used, in synergy with magnetohydrodynamic simulations, to study the imprint of turbulence and magnetization on the formation of structure in the cloud \citep[][]{Houde2009,Soler2013}.




Ground-based observations with the SCUBA polarimeter
\citep{Murray1997,Greaves2003} and the Submillimeter Polarimeter
for Antarctic Remote Observations \citep[SPARO;][]{Novak2003} show
that the submm emission from prestellar cores and giant molecular clouds (GMCs) is indeed
polarised to a few percent \citep{WardThompson2000,Li2006}. {\sl
Planck} \citep{Planck2011} will provide coarse resolution ($\sim$5\arcmin) submm polarimetry maps of the entire Galaxy. The
Atacama Large Millimeter/submillimeter Array
\citep[ALMA;][]{Wootten2009} will provide sub-arcsecond millimetre (mm) and submm polarimetry, capable of resolving fields
within cores and circumstellar disks, but will not be sensitive to
cloud-scale fields.

The Balloon-borne Large-Aperture Submillimeter Telescope for
Polarimetry
\citep[BLASTPol;][]{Marsden2008,Fissel2010,Pascale2012}, with its
arcminute resolution, is the first submm polarimeter to map the
large-scale magnetic fields within molecular clouds with unique combined sensitivity and mapping speed, and sufficient angular resolution
to observe into the dense cores. BLASTPol will be able to trace magnetic structures in
the cold interstellar medium from scales of 0.05\,pc out to 5\,pc, thus providing
a much needed bridge between the large-area but coarse-resolution
polarimetry provided by {\sl Planck} and the high-resolution but
limited field-of-view maps of ALMA.


BLASTPol successfully completed two science flights over Antarctica during the austral summers of 2010 and 2012, mapping the polarised dust emission at 250, 350, and 500\,$\mu$m over a wide range of column densities corresponding to $A_V$\,$\gtrsim$\,4\,mag, yielding hundreds to thousands of independent polarisation pseudo-vectors per cloud, for a dozen between GMCs and dark clouds. The first scientific results from the 2010 campaign are soon to be released \citep[][and Poidevin et al. in preparation]{Matthews2013}, while the 2012 data are still under analysis (an overview of the 2012 observations can be found in Angil\`{e} et al. in preparation).


The BLASTPol linear polarisation modulation scheme comprises a
stepped cryogenic achromatic half-wave plate (HWP) and
photolithographed polarising grids placed in front of the detector
arrays, acting as analysers. The grids are patterned to alternate
the polarisation angle sampled by 90$^{\circ}$ from
bolometer-to-bolometer along the scan direction. BLASTPol scans so
that a source on the sky passes along a row of detectors, and thus
the time required to measure one Stokes parameter (either $Q$ or
$U$) is just equal to the separation between bolometers divided by
the scan speed. During operations, we carry out spatial scans at
four HWP rotation angles spanning 90$^{\circ}$ (22.5$^{\circ}$
steps), allowing us to measure the other Stokes parameter through
polarisation rotation.

The use of a continuously rotating or stepped HWP as a
polarisation modulator is a widespread technique at (sub)mm
wavelengths
\citep[e.g.,][]{Renbarger2004,Hanany2005,Pisano2006,Savini2006,Savini2009,Johnson2007,Li2008a,Matsumura2009}.
A thorough account of the HWP non-idealities and its inherent
polarisation systematics, especially for very achromatic designs,
has become necessary as the accuracy and sensitivity of (sub)mm
instruments have soared in recent years.

The literature offers numerous efforts to address, through
simulations, the impact of the inevitable instrumental systematic
errors due to the polarisation modulation strategy in the unbiased
recovery of the Stokes parameters $Q,U$ on the sky, especially for
cosmic microwave background (CMB) polarisation experiments
\citep[e.g.,][]{Odea2007,Odea2011,Brown2009}. In addition,
physical and analytical models have been developed to retrieve the
frequency-dependent modulation function of achromatic HWPs and
estimate the corrections due to non-flat source spectral indices
\citep{Savini2006,Savini2009,Matsumura2009}.

Nevertheless, little work has been published on incorporating the
{\it measured} HWP non-idealities in a data-analysis pipeline and
ultimately in a map-making algorithm. \citet{Bryan2010} derive an
analytic model that parametrises the frequency-dependent
non-idealities of a monochromatic HWP and present a map-making
algorithm that accounts for these. \citet{Bao2012}
carefully simulate the impact of the spectral dependence of the
polarisation modulation induced by an achromatic HWP on
measurements of the CMB polarisation in the presence of
astrophysical foregrounds, such as Galactic dust. However, both
these works assume the nominal design values for the build
parameters of the HWP plus anti-reflection coating (ARC) assembly.

While this assumption is a reasonable one when no spectral
measurements of the HWP as-built are available, several studies
clearly show that the complex multi-slab crystal HWP and its
typically multi-layer ARC are practically impossible to
manufacture {\it exactly} to the desired specifications. In
particular, \citet{Savini2006,Savini2009} and \citet{Pisano2006}
caution against the finite precision to which the multiple crystal
substrates composing an achromatic HWP can be aligned relative to
each other in the Pancharatnam \citeyearpar{Pancharatnam1955}
scheme. In addition, \citet{Zhang2009} show how some of the design
parameters in the ARC can slightly change during the bonding of
layers, achieved via a hot-pressing technique \citep{Ade2006}. We will briefly cover these points and discuss the repercussions on the HWP performance.

This work describes a novel empirical method that allows the
reconstruction of the Mueller matrix\footnote{We adopt the Stokes
\citeyearpar{Stokes1852} formalism to represent the time-averaged
polarisation state of electromagnetic radiation; for a review of
polarisation basics we refer the reader to \citet{Collett1993}.}
of a generic HWP as a function of frequency through spectral
transmission measurements of the HWP rotated by different angles
with respect to the input polarised light. Not only does this
method give complete and quantitative information on the {\it
measured} spectral performance of the HWP, but it also provides a
direct avenue to accounting for the non-idealities of the HWP {\it
as-built} in a map-making algorithm. This empirical approach is
applied to the BLASTPol HWP and will help improve the accuracy on astronomical measurements of
polarisation angles at submm wavelengths.


The layout of this paper is as follows. In
Section~\ref{sec:intro_HWP}, we give an overview of the
manufacturing process for BLASTPol's five-slab sapphire HWP.
Section~\ref{sec:HWP_spectra} describes the spectral measurements,
while Section~\ref{sec:empirical_model} presents the empirical
model as well as the main results of the paper. Finally, in
Section~\ref{sec:map_maker}, we describe the algorithm for the
naive-binning map-making technique implemented by BLASTPol, which
naturally accounts for the {\it measured} HWP non-idealities.
Section~\ref{sec:concl_HWP} contains our conclusions.

\section{The BLASTP\lowercase{ol} Half-Wave Plate}\label{sec:intro_HWP}

Wave plates (or retarders) are optical elements used to change the
polarisation state of an incident wave, by inducing a
predetermined phase difference between two perpendicular
polarisation components. A (monochromatic) wave plate can be
simply obtained with a single slab of uniaxial birefringent
crystal of specific thickness, which depends upon the wavelength
and the index of refraction of the crystal. A birefringent crystal
is characterised by four parameters, $n_{\rm e}$, $n_{\rm o}$,
$\alpha_{\rm e}$, $\alpha_{\rm o}$, the real part of the indices
of refraction and the absorption coefficient (in cm$^{-1}$) for
the extraordinary and ordinary axes of the crystal. At a specific
wavelength $\lambda_0$, the phase shift induced by a slab is
determined uniquely by its thickness $d$, and reads:
\begin{equation}\label{eq:phase_HWP}
\Delta\varphi\left(\lambda_0\right) = \frac{2\,\pi\,d}{\lambda_0}
\left(n_{\rm e} - n_{\rm o}\right)~.
\end{equation}%
Given the operating wavelength $\lambda_0$, the required phase
shift for the wave plate is achieved by tuning the thickness $d$.

While monochromatic wave plates have been (and are still being)
used in (sub)mm astronomical polarimeters \citep[see
e.g.,][]{Renbarger2004,Li2008a,BryanSPIE,Bryan2010,Dowell2010},
the inherent dependence of the phase shift on wavelength,
expressed in Equation~\ref{eq:phase_HWP}, constitutes an intrinsic
limit in designing a polarisation modulator that operates in a
broad spectral range (i.e., is achromatic).

\subsection{Achromatic half-wave plate design}\label{sec:HWP_design}

Achromaticity is necessary for wave plates that are designed for
use with multi-band bolometric receivers, such as BLASTPol, PILOT
\citep{Bernard2007}, or SCUBA-2 \citep{Bastien2005,Savini2009}. To
achieve a broadband performance, multiple-slab solutions have been
conceived in the past \citep{Pancharatnam1955,Title1981} to
compensate and keep the phase shift approximately constant across
the bandwidth, by stacking an odd number (usually 3 or 5) of
birefringent substrates of the same material, which are rotated
with respect to each other about their optical\footnote{We
distinguish between ``optic'' axis of a crystal, i.e. the
direction in which a ray of transmitted light experiences no
birefringence, and ``optical'' axis, i.e. the imaginary line along
which there is some degree of rotational symmetry in the optical
system.} axes by a frequency-dependent set of angles.

Achromatic wave plates have been designed and built for
astronomical polarimeters at (sub)mm wavelengths by many
authors in the last decade
\citep{Hanany2005,Pisano2006,Savini2006,Savini2009,Matsumura2009},
following the Poincar\'{e} sphere (PS) method first introduced by
Pancharatnam \citeyearpar{Pancharatnam1955}. Because the four parameters that characterise a crystal, $n_{\rm e}$, $n_{\rm o}$, $\alpha_{\rm
e}$, $\alpha_{\rm o}$, all depend upon wavelength (in particular, the different
frequency-dependence of the ordinary and extraordinary refraction
indices enters Equation~\ref{eq:phase_HWP} in a non-trivial way, as we will
illustrate in detail for sapphire), the design of an achromatic HWP becomes progressively more difficult as the bandwidth increases.

\begin{figure}
\centering
\includegraphics[width=1.\linewidth]{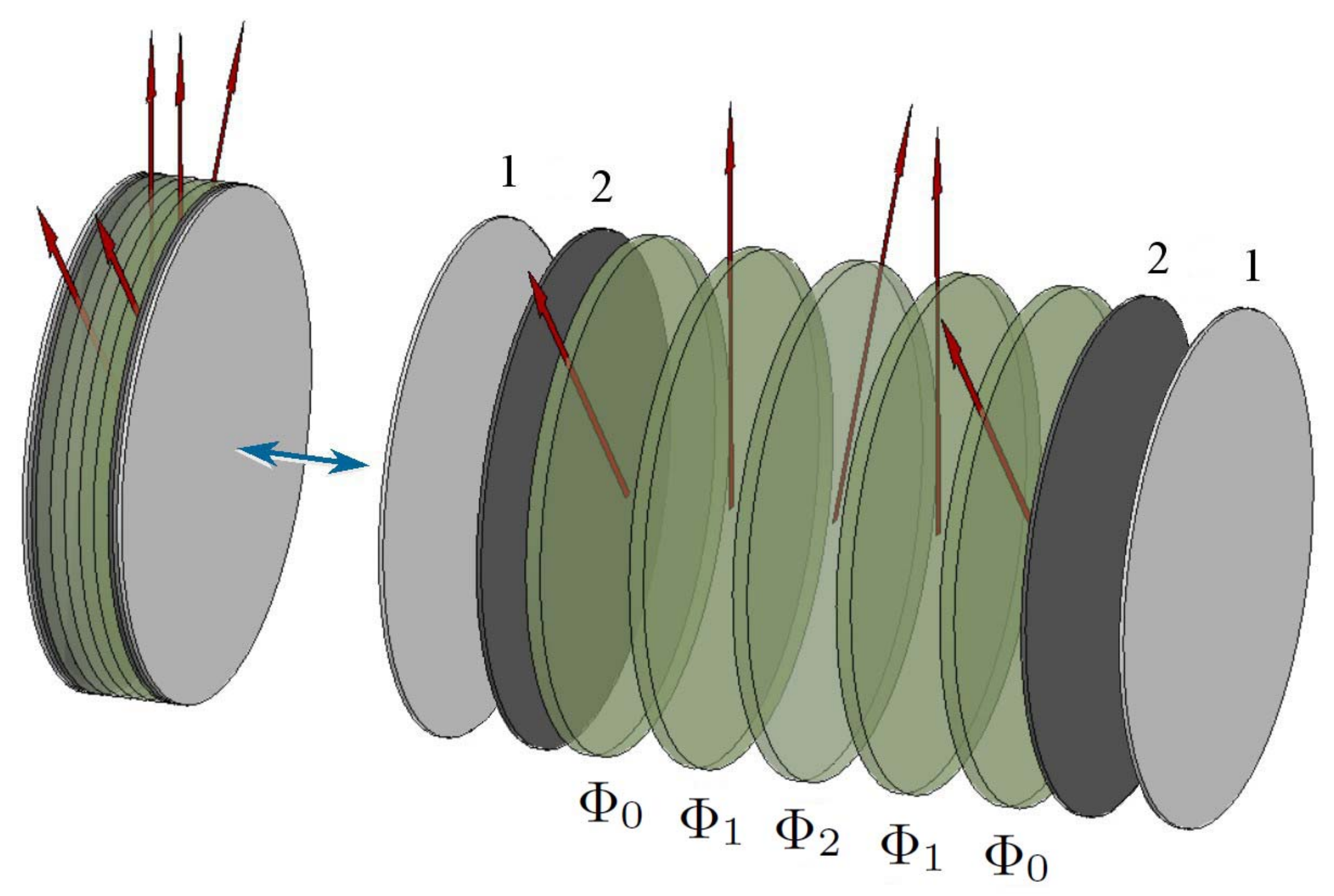}
\caption{Exploded view of the BLASTPol HWP. We also show the
two-layer anti-reflection coating described in
Section~\ref{sec:HWP_manu}. \citep[Figure modified from][]{Savini2006}.}\label{fig:HWP_exploded_plates}
\end{figure}

Using the PS method, we have designed a HWP for the BLASTPol
instrument, which requires an extended frequency range to cover
three adjacent 30\% wide spectral bands at 250, 350, and
500\,$\mu$m. A Pancharatnam \citeyearpar{Pancharatnam1955}
five-slab design is chosen with axis orientations of
$\Phi_0=0^{\circ}$, $\Phi_1=26^{\circ}$, $\Phi_2=90.3^{\circ}$,
$\Phi_1=26^{\circ}$, and $\Phi_0=0^{\circ}$; these angles are
optimised using the physical and analytical model developed by
\citet{Savini2006} for an achromatic HWP, which in turn is based
on the work of \citet{Kennaugh1960}. In
Fig.~\ref{fig:HWP_exploded_plates} we show an exploded view of the HWP assembly; to our knowledge, with its $\sim$100\%
bandwidth, this is the broadest-band HWP
manufactured to date at (sub)mm wavelengths for which measurements are published\footnote{The HWP
designed for the mm-wave E and B EXperiment
\citep[EBEX;][]{Matsumura2009,Reichborn2010} is nominally slightly
more achromatic, with a $\sim$110\% bandwidth. However, a
comprehensive spectral characterisation of the final (as-flown)
AR-coated HWP assembly has yet to be published.}.

\subsection{HWP manufacture}\label{sec:HWP_manu}

In addition to the broad spectral range of operation, the BLASTPol
HWP is required to function at cryogenic temperatures \citep[4\,K;
see][]{Fissel2010} for two main reasons: (1) reduce the thermal
emission from a warm element placed in the optical path, which
would constitute a significant background load on the bolometers;
and (2) reduce the losses in transmission due to absorption from
the stack of five crystal substrates, which drops dramatically
with temperature. The absorption in a crystal at FIR wavelengths
is the result of the interactive coupling between the incident
radiation and phonons -- the thermally induced vibrations of the
constituent atoms of the substrate crystal lattice. Because the
phonon population is much reduced at low temperatures, cooling the
crystal effectively reduces the absorption.

The two obvious candidates (uniaxial birefringent) crystals are
sapphire and quartz, because of their favorable optical properties
in the FIR/submm. Sapphire is chosen over quartz due to its larger
difference between ordinary and extraordinary refraction indices
\citep[$\Delta n_{\rm e-o} \approx 0.32$ for sapphire, and
$\approx 0.032$ for quartz;][]{Loewenstein1973}, which implies a
smaller thickness for the substrates (see
Equation~\ref{eq:phase_HWP}). Since quartz and sapphire have a
comparable level of absorption at cryogenic temperatures in the
wavelength range of 200--600\,$\mu$m \citep{Loewenstein1973},
thinner substrates are desirable to minimise absorption losses.
Nonetheless, the thin sapphire substrates chosen for the BLASTPol
HWP do indeed show appreciable absorption, especially at the
shortest wavelengths (250\,$\mu$m band; see
Section~\ref{sec:HWP_spectra}).

The five slabs of the Pancharatnam \citeyearpar{Pancharatnam1955}
design all have the same thickness. To cover the broad wavelength
range of 200--600\,$\mu$m, a substrate thickness is chosen to
produce a HWP at the central wavelength of the central band,
350\,$\mu$m. By using the values of the refractive indices for
cold sapphire published by \citet{Loewenstein1973} and
\citet[][$\Delta n^{\rm 350\,\mu m}_{\rm e-o} \approx
0.32$]{Cook1985}, and imposing the required phase shift of
180$^{\circ}$ between the two orthogonal polarisations travelling
through the plate, Equation~\ref{eq:phase_HWP} yields for the
thickness of a single substrate a value $\sim$0.547\,mm. The
nearest available thickness on the market is 0.5\,mm, and sapphire can not be easily ground to the desired thickness due to its brittleness. A deviation
of $\sim$0.047\,mm from the desired thickness would naively translate into a departure of up to $\sim$15$^{\circ}$ from the ideal phase shift of
180$^{\circ}$ at 350\,$\mu$m. However, the case of a multi-slab Pancharatnam HWP is more complex, since the phase shift becomes an ``effective'' one and requires proper modelling as a function of frequency, which we have included in Section~\ref{sec:phase_shift}.

The orientation of the optic axis on each sapphire substrate is
determined with a polarising Fourier transform spectrometer (pFTS
hereafter), which is briefly described in Section~\ref{sec:cold_spectra}.
Each substrate is rotated between two aligned polarisers at the
pFTS output until a maximum signal is achieved. The use of two
polarisers avoids any complication from a partially polarised
detecting system and any cross polarisation incurred from the pFTS
output mirrors. The HWP is assembled by marking the side of each
substrate with its reference optic axis and rotating each element
according to the Pancharatnam design described in the previous
section. The stack of five carefully-oriented sapphire substrates,
interspersed with 6\,$\mu$m layers of polyethylene, are bonded
together with a hot-pressing technique used in standard FIR/submm
filter production \citep{Ade2006}. The polyethylene has negligible
effects on the final optical performance of the HWP, because when
heated it seeps into the roughened surfaces of the adjacent
substrates.

In order to improve the robustness of the bond, the individual
substrates are sandblasted with aluminium oxide (Al$_2$O$_3$)
prior to fusion; this procedure dramatically improves the grip of
the polyethylene between adjacent crystal surfaces. Careful
cleansing and degreasing of all the crystal surfaces is required
after sandblasting; in particular, we found trichloroethylene to
be most effective in removing the traces of oily substances due to
the sandblasting process.

A two-layer
broadband anti-reflection coating (ARC), necessary to maximise the
in-band transmission of the HWP, is also hot-pressed to the front
and back surfaces of the assembled plate, again using 6\,$\mu$m
layers of polyethylene. The layer adjacent to the sapphire is an
artificial dielectric metamaterial (ADM) composed of metal-mesh
patterned onto polypropylene sheets \citep{Zhang2009}, while the
outer layer is a thin film of porous polytetrafluoroethylene (PTFE). The
thickness of the final stack (coated HWP) is $2.80\pm 0.01$\,mm.
The diameter of the ARC is set to $88.0\pm 0.1$\,mm, slightly
smaller than that of the HWP ($100.0\pm 0.1$\,mm) to avoid any contact between the
coating and the HWP mount \citep[see][]{Fissel2010}; the ARC is
bonded concentrically to the HWP and thus its diameter defines the
optically-active area of the HWP.

Because of the thermal expansion mismatch between the sapphire and
the polypropylene, the HWP assembly has been cryogenically cycled
numerous times prior to the flight to test the robustness of the
bond at liquid helium temperatures. The HWP has been successfully
installed in the BLASTPol cryogenic receiver and has been flown twice from a
balloon platform, without delamination of the
ARC or damage to the assembly.

However, for cryogenic crystal HWPs much larger than the
BLASTPol one, the application of a metal-mesh ADM as an ARC has
proven extremely challenging. Therefore, extending previous work
by \citet{Pisano2008}, we have recently designed and realised a
prototype polypropylene-embedded metal-mesh broadband achromatic
HWP for mm wavelengths \citep{Zhang2011}; this will allow next
generation experiments with large-aperture detector arrays to be
equipped with large-format ($\gtrsim$\,20\,cm in diameter) HWPs
for broadband polarisation modulation.

\section{Spectral characterisation}\label{sec:HWP_spectra}

The first step to retrieving the frequency-dependent Mueller matrix and phase shift of the BLASTPol HWP is to measure its transmission as a function of frequency and incoming polarisation state. Because of the strong dependence of the sapphire absorption coefficient on temperature, we can not limit ourselves to determining the room-temperature response of the HWP, which is designed to operate at cryogenic temperatures. Therefore, we measure its spectral response in a vacuum cavity, cooled to temperatures as low as currently possible with our experimental apparatus ($\sim$120\,K)\footnote{Room-temperature spectra were acquired and can be made available to the reader, but are neither reported here nor used in the subsequent analysis, because plagued by significant in-band transmission loss due to the absorption from sapphire.}.

\subsection{Experimental setup}\label{sec:cold_spectra}

\begin{figure}
\centering
\includegraphics[width=1.\linewidth]{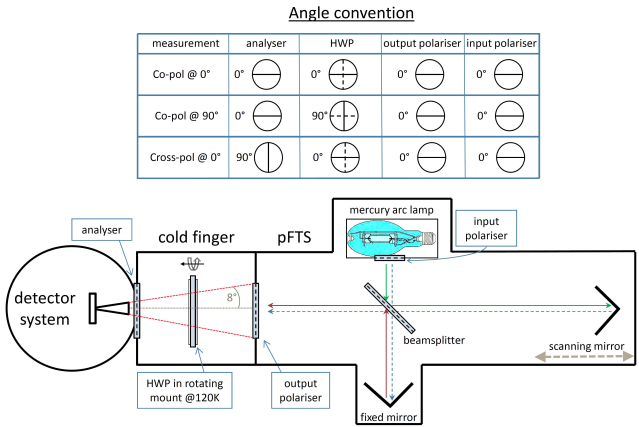}
\caption{Schematic drawing of the spectral
measurements setup. The $f/3.5$ horizontally polarised output of a pFTS illuminates the HWP with a $\sim$8$^{\circ}$ incidence angle and is focused directly onto the horn aperture of the bolometric
detector. The analyser alternatively parallel or perpendicular to the pFTS output polariser creates the necessary polarisation selection for the ``co-pol'' and
``cross-pol'' sets of measurements, as depicted in the table above. The lines in the circles indicate the selected polarisation, so that for the photolithographed polarisers the wire grid orientation is perpendicular to the lines.}\label{fig:FTS_scheme_HWP_measurements}
\end{figure}

We fully characterise the spectral performance of the BLASTPol HWP
by using a pFTS of the Martin-Puplett
\citeyearpar{MartinPuplett1970} type. A schematic drawing of the experimental setup is shown in Fig.~\ref{fig:FTS_scheme_HWP_measurements}; in the following, we describe each element in sequential order from the source to the detector system.

The source is an incoherent
mercury arc lamp with an aperture of 10\,mm, whose emission is
well approximated by a blackbody spectrum at $T_{\rm eff}\approx
2000$\,K; a low-pass filter blocks radiation from the source at
wavelengths shorter than $\sim$3.4\,$\mu$m. The interferometer is
equipped with a P1\footnote{The notation P{\it n} refers to a wire grid polariser that has a grid period of {\it n}\,[$\mu$m], with {\it
n}/2 copper strips and {\it n}/2 gaps, photolithographed on a 1.5\,$\mu$m mylar
substrate.} beam divider, a P2 input polariser (at the source),
and a P10 output polariser. The pFTS has a (horizontally)
polarised output focused beam with $f/3.5$ or, in other words, a
converging beam with angles $\theta\la8^{\circ}$.

This beam spread is conveniently close to the $\sim$5.7$^{\circ}$ incidence angle that the HWP is illuminated by in the $f/5$ BLASTPol optics box \citep[see][]{Fissel2010,Pascale2012}, therefore it is ideal optically to place the HWP between the pFTS output polariser and the detector system the beam focuses onto, without the need for additional optical elements; this also ensures an even illumination of the entire HWP optically-active area.

\begin{figure}
\centering
\includegraphics[width=1.\linewidth]{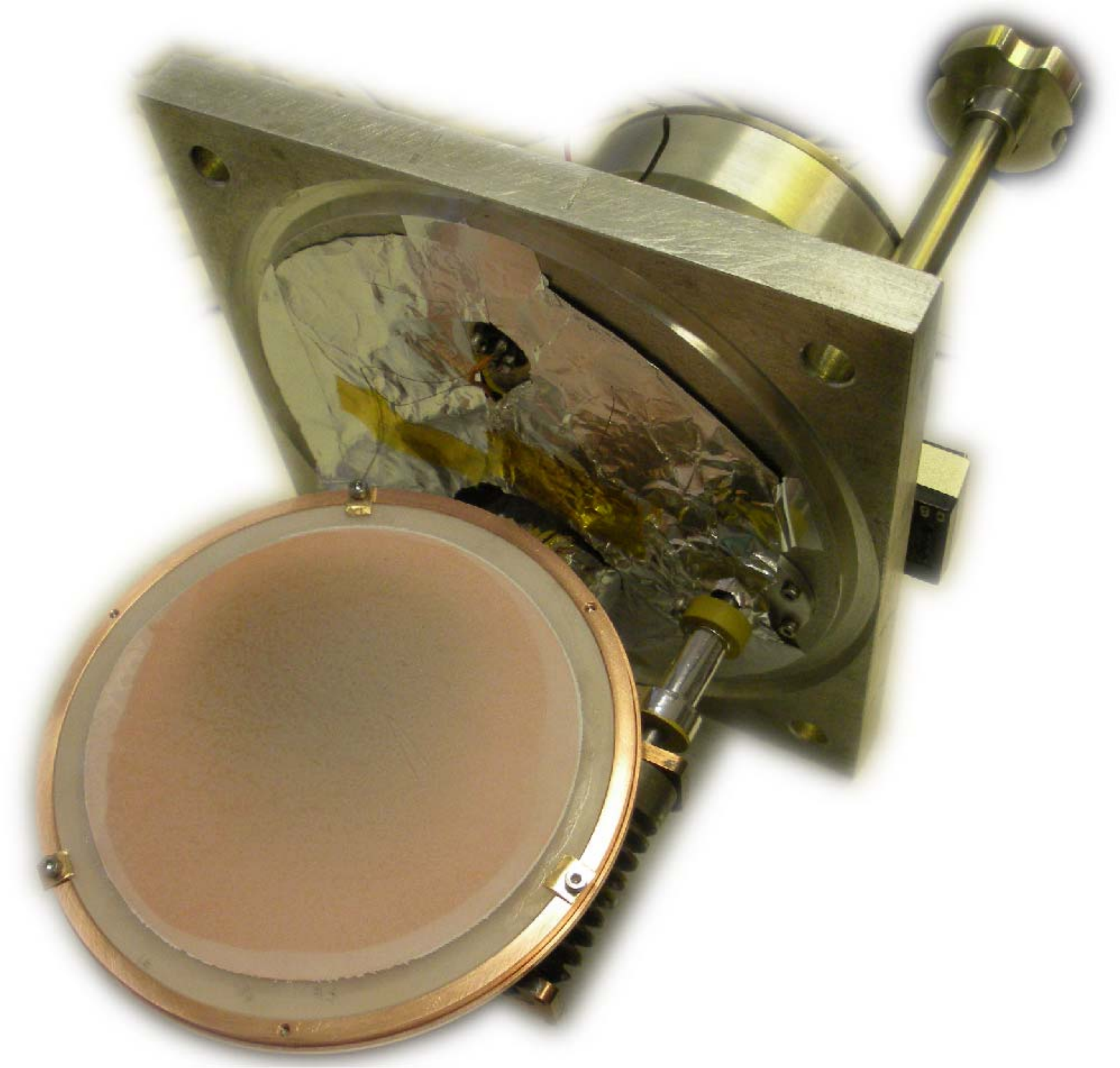}
\caption{Photograph of the ``cold finger'' module, which fits in
the vacuum cavity at the output port of the pFTS. The central
cylinder is hollow and has to be regularly replenished with liquid
nitrogen to maintain the temperature of the HWP at $\sim$120\,K.
Aluminium insulation and a thick copper strap improve the thermal
performance of the module. Two thermometers monitor the
temperature at the bottom of the cylinder (base plate) at the edge
of the copper HWP holder. The rotator is manually driven via a
gear train and a vacuum-seal shaft leading to a manual knob
outside the module. The resolution of the analogue encoder on the
rotation angle is 0.06$^{\circ}$. The presence of a thermometer on
the rotating element prevents rotations greater than
$\sim$180$^{\circ}$.} \label{fig:cold_finger}
\end{figure}

We position the HWP in a liquid nitrogen-cooled removable module
retrofitted in the vacuum cavity at the output port of the pFTS; a
photograph and a brief description of the module, which we refer
to as ``cold finger'', are given in Fig.~\ref{fig:cold_finger}. The manually-driven rotating mount can rotate the HWP about
its optical axis to obtain the polarisation modulation, with a resolution on the rotation angle of 0.06$^{\circ}$.

The HWP is placed centrally between the pFTS output polariser and a P10 analyser, installed at the
exit port of the vacuum cavity; the efficiency of these polarisers is
separately determined to exceed 99.8\% over the range of
frequencies of interest, with a cross-polarisation of less than
0.1\%. The polarisers are initially aligned with respect to each
other, with the grid wires vertical (thus selecting horizontal
polarisation) with respect to the optical bench.

A small cryostat, connected with no air gaps to the exit port of the vacuum cavity, houses a feedhorn-coupled composite bolometer cooled to 1.5\,K by pumping on the liquid helium bath. The spectral coverage of the data is thus defined by the cut-off frequency of the light collector waveguide (5\,cm$^{-1}$) and by a low-pass filter (60\,cm$^{-1}$) installed in the cryostat to minimise photon noise.

Finally, the rapid-scan system records interferograms with a 8\,$\mu$m
sampling interval over a 10\,cm optical path difference, at a scan
speed of 2\,cm\,s$^{-1}$; this results in a Nyquist frequency of
625\,cm$^{-1}$ and a spectral resolution of 0.05\,cm$^{-1}$.

\subsection{Measurement strategy and results}\label{sec:measurements_and_results}

After the roughly two hours needed for the cold finger module to
thermalise, its base plate reaches temperatures close to 77\,K,
while the HWP holder thermalises at about 120\,K, despite the thermal insulation and high thermal conductivity link to the base plate. Other
cryogenic tests conducted by bonding a thermometer at the center
of a single slab of sapphire ensure that the temperature measured
at the edge of an aluminium or copper holder closely matches that
of the sapphire substrate at its center. While maintaining a constant level of liquid nitrogen in the cold
finger, we can characterise the spectral response of the cold HWP,
by rotating it inside the vacuum cavity.

Following the convention depicted in
Fig.~\ref{fig:FTS_scheme_HWP_measurements}, measurements with
aligned polarisers are referred to as ``co-pol'' transmission,
$T_{\rm cp}$. The HWP has a
complementary response when the analyser is
rotated by 90$^{\circ}$ about the optical axis of the system
(i.e., horizontal wires, selecting vertical polarisation); data
taken with this configuration are necessary to completely
characterise the HWP, and are referred to as ``cross-pol''
transmission\footnote{We note that this definition of cross-pol
may differ from other conventions adopted in the literature (e.g.,
that of \citealt{Masi2006}, who operate without a HWP).}, $T_{\rm
xp}$.

The very first dataset, which we refer to as
the background spectrum, must be obtained in co-pol configuration by scanning
the spectrometer in the absence of the HWP. This dataset defines the pFTS reference
spectral envelope, and it is the set against which all the
following spectra are divided in order to account for the spectral
features of the source, pFTS optics, and detector system. Subsequently, the HWP is inserted in between the polarisers in co-pol
configuration, and spectra are acquired at many different HWP
rotation angles (resulting in a data cube). To enhance the
spectral signal-to-noise ratio, each dataset at a given angle is
obtained by computing the Fourier transform of an (apodised and
phase-corrected) average of 60 interferograms with the mirror
scanned in both the forward and backward directions. As
anticipated, the resulting spectra are divided by the background
dataset, which in turn is the average of three spectra, to obtain
the transmission of the coated HWP alone as a function of
frequency.

\begin{figure}
\centering
\includegraphics[width=1.\linewidth]{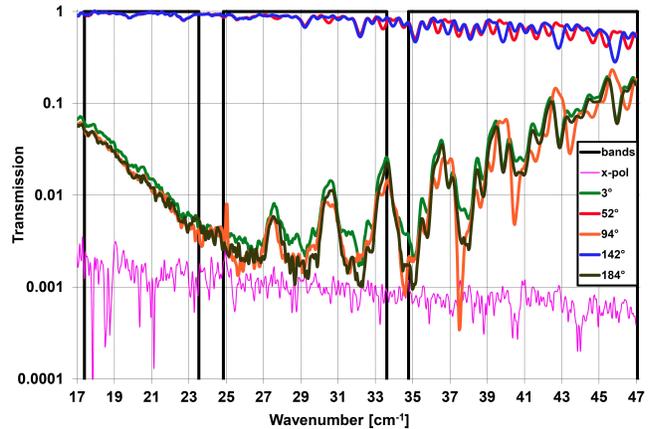}
\caption{Measured co-pol transmission spectra of the coated
BLASTPol HWP cooled to $\sim$120\,K. Each line is obtained at a
different HWP rotation angle by computing the Fourier transform of
an (apodised and phase-corrected) average of 60 interferograms. For visual clarity, we only show here spectra at rotation angles near the HWP maxima and minima, omitting data taken at intermediate angles (shown in Fig.~\ref{fig:spareHWP_copol_spectra_surface}). The solid vertical black lines show
the approximate extent of the three BLASTPol bands.} \label{fig:BLAST_pol_spareHWP_ARC_copol_spectra}
\end{figure}

\begin{figure}
\centering
\includegraphics[width=1.\linewidth]{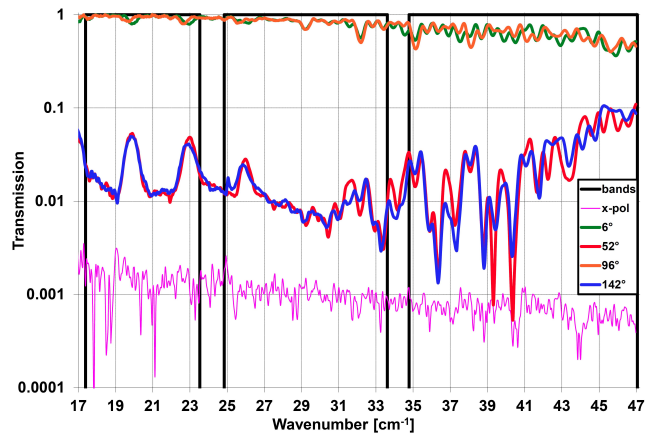}
\caption{Measured spectra of the HWP cooled to $\sim$120\,K equivalent to
those shown in Fig.~\ref{fig:BLAST_pol_spareHWP_ARC_copol_spectra}
but for cross-pol transmission.}
\label{fig:BLAST_pol_spareHWP_ARC_xpol_spectra}
\end{figure}

\begin{figure}
\centering
\includegraphics[width=1.\linewidth]{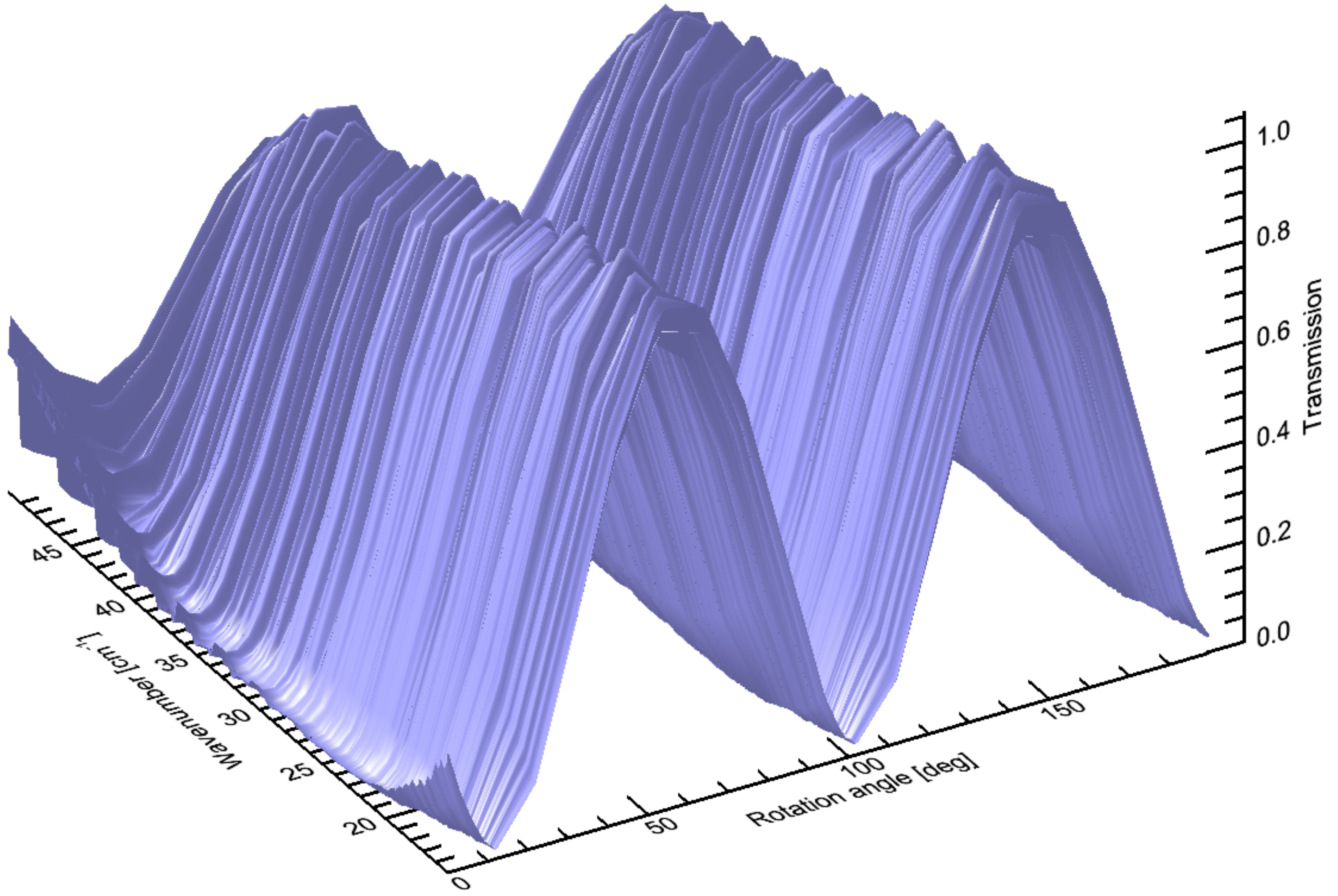}
\caption{Data cube represented by a surface obtained by stacking a
set of spectral co-pol transmissions of the HWP at different
angles. Each measured spectra is a slice of
the surface perpendicular to the angle axis.}
\label{fig:spareHWP_copol_spectra_surface}
\end{figure}

\begin{figure}
\centering
\includegraphics[width=1.\linewidth]{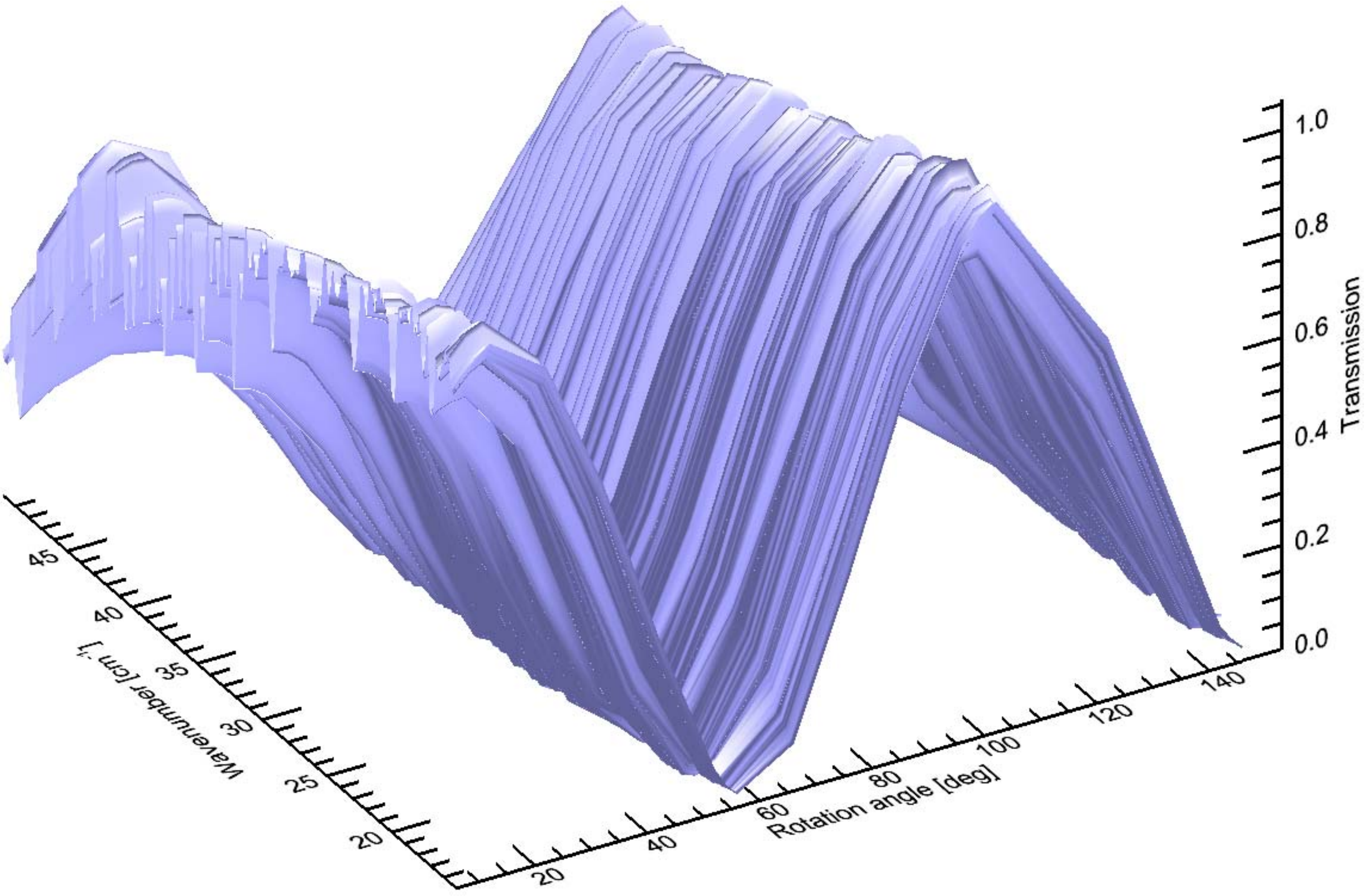}
\caption{Equivalent data cube to that shown in
Fig.~\ref{fig:spareHWP_copol_spectra_surface} but for cross-pol
transmission. Note how the two surfaces are complementarily in
counterphase to each other. Each measured spectra is a slice of
the surface perpendicular to the angle axis.}
\label{fig:spareHWP_xpol_spectra_surface}
\end{figure}

Over two days of measurements, we acquire data cubes for co-pol and cross-pol transmissions, shown respectively in Fig.~\ref{fig:BLAST_pol_spareHWP_ARC_copol_spectra} and Fig.~\ref{fig:BLAST_pol_spareHWP_ARC_xpol_spectra}, where we only display spectra taken at rotation angles near the HWP maxima and minima, for visual clarity. The full datasets, including spectra taken at intermediate angles (roughly every 10$^{\circ}$ between 0$^{\circ}$ and $\sim$180$^{\circ}$), are shown in Figs.~\ref{fig:spareHWP_copol_spectra_surface} and \ref{fig:spareHWP_xpol_spectra_surface} as 3-D surfaces.

Because of the controlled environment in the vacuum
cavity, our measurements are not susceptible to changes in the external environment; however, we repeat background scans at the
very end of our measurement session to monitor drifts in the
bolometer responsivity and other potential systematic effects.
Prior to inserting the HWP in the cavity, we have also characterised
the instrumental cross-pol of this setup by rotating the analyser by
90$^{\circ}$ in cross-pol configuration and acquiring three
spectra. By averaging these cross-pol spectra and dividing by the
co-pol background, we measure a cross-pol level of 0.2\% or less
across the entire spectral range of interest (5--60\,cm$^{-1}$);
we include the resulting cross-pol spectrum in
Figs.~\ref{fig:BLAST_pol_spareHWP_ARC_copol_spectra} and
\ref{fig:BLAST_pol_spareHWP_ARC_xpol_spectra} (dark pink line).

An ideal HWP modulates the polarisation at 4\,$\theta$, therefore
in a complete revolution there are four maxima (and minima), two
for each of the birefringent axes. The (arbitrary) zero angle in Figs.~\ref{fig:BLAST_pol_spareHWP_ARC_copol_spectra}, \ref{fig:BLAST_pol_spareHWP_ARC_xpol_spectra}, \ref{fig:spareHWP_copol_spectra_surface}, and \ref{fig:spareHWP_xpol_spectra_surface} does not (need to) coincide with a HWP maximum (minimum), which is the HWP angle at which we measure maximum (minimum) total power on the detector; this of course
includes signal outside of the HWP bands (in the range
5--60\,cm$^{-1}$). As we show in Section~\ref{sec:beta_ea}, the
position of the equivalent axes of the sapphire plate stack (and
hence the position of the HWP maxima/minima) depends upon the
wavelength. Therefore the HWP maxima (and minima) we assign while
taking spectra are just rough approximations. Although we increase
the angle sampling rate in the vicinities of a maximum or minimum (with steps of 3$^{\circ}$ rather than 10$^{\circ}$),
in order to fully characterise the HWP it is not necessary to take
spectra {\it exactly} at its maxima or minima.

We verify that the experimental setup is symmetric with respect to the HWP rotation and that there are no artifacts arising from misalignments in the optical setup by measuring no appreciable change in pairs of datasets taken at angles that are exactly 180$^{\circ}$ apart, due to polarisation symmetry.


In the surfaces depicted in
Figs.~\ref{fig:spareHWP_copol_spectra_surface}
and~\ref{fig:spareHWP_xpol_spectra_surface}, slices of the data
cube along the wavenumber axis constitute the measured spectra at
different HWP angles, while slices along the angle axis
represent the modulation function of the HWP at a given
frequency or, more precisely, within a narrow band of frequencies
defined by a combination of spectral resolution and the
spectrometer's instrument response function.

The features visible in all spectra are
spectral fringes due to standing waves generated inside the stack
of dielectric substrates (even with a quasi-perfect impedance
matching coating on the outer surfaces); the presence of several
interspersed layers of polyethylene enhances the amplitude of the
fringes by introducing small amounts of absorption at every
internal reflection.

\subsection{Uncertainties on the measured spectra}\label{sec:uncertainties _spectra}

\begin{figure}
\centering
\includegraphics[width=1.\linewidth]{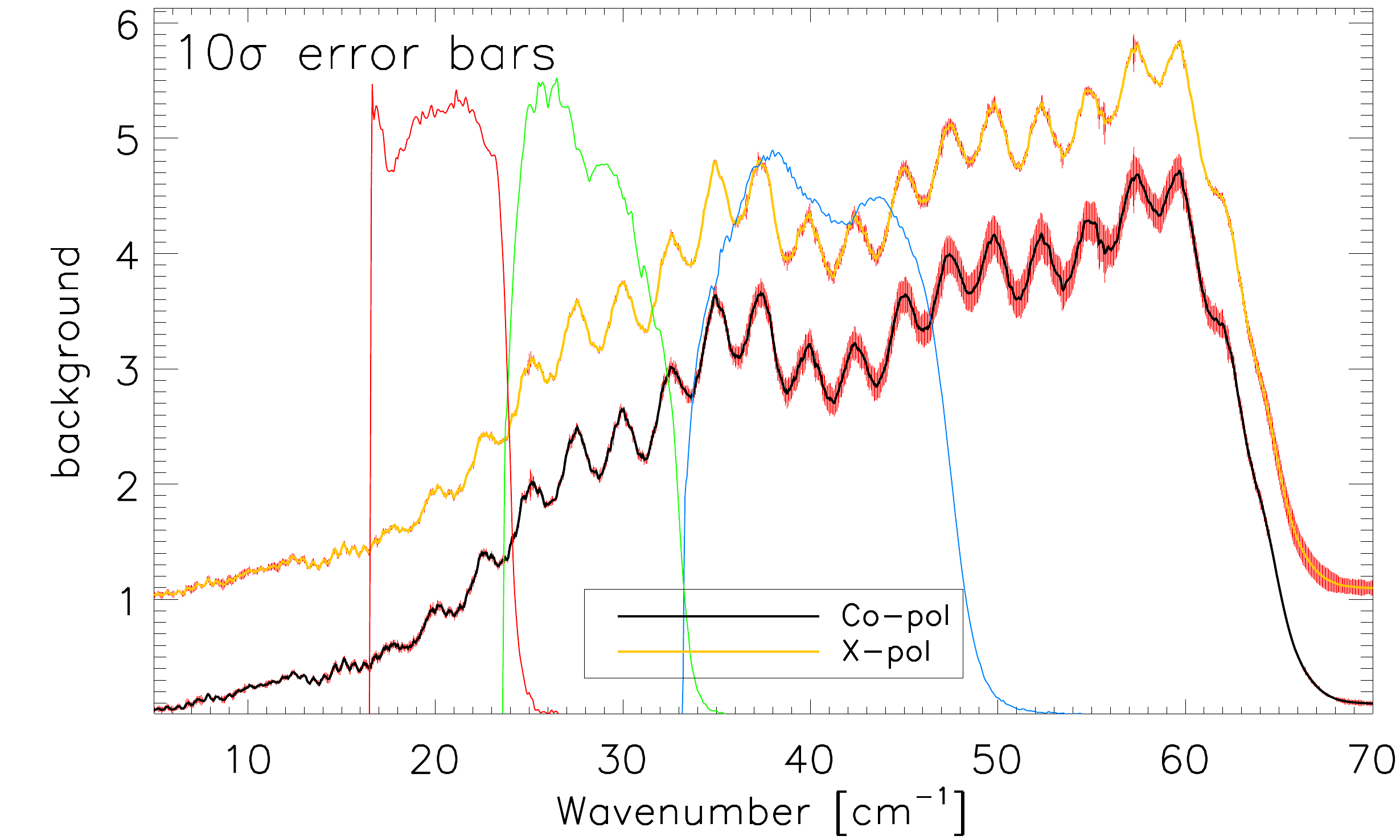}
\caption{Noise estimation for the spectra shown in
Figs.~\ref{fig:BLAST_pol_spareHWP_ARC_copol_spectra} and
\ref{fig:BLAST_pol_spareHWP_ARC_xpol_spectra}. We plot the mean
background spectra (in arbitrary units) for the co-pol (black
solid line) and cross-pol (yellow solid line, shifted by 1 in the
positive $y$ direction for visual clarity) as a function of
wavenumber. The (10\,$\sigma$) error bars (in red) are quantified
as the statistical error on the mean. Also shown for reference is
the relative spectral response of the three BLASTPol channels, in
arbitrary units. Henceforth, we adopt a colour code in the plots
whereby curves referring to the three BLASTPol bands, 250, 350,
and 500\,$\mu$m are drawn in blue, green, and red, respectively.}
\label{fig:plot_noise_backs_HWP}
\end{figure}

Because we average 60 interferograms to obtain the final spectrum
at each HWP position, the statistical uncertainty associated with
the average on a single dataset is found to be negligible, as
expected. Rather, we decide to average together all the available
background interferograms that are collected over one day of
measurements, and take their statistical dispersion as our
estimate of the uncertainty associated with all the spectra
collected on that day. Because the thermodynamic conditions in the
cavity under vacuum are not susceptible to changes in the external
environment, this procedure allows us to account for drifts in the
bolometer responsivity and other potential systematic effects. We report in Fig.~\ref{fig:plot_noise_backs_HWP} the mean background
spectra and the associated error (shown as 10\,$\sigma$ error bars for visual clarity) for the co-pol and cross-pol
measurement sessions. These errors (1\,$\sigma$) are used in
Section~\ref{sec:empirical_model} to estimate the uncertainties on
the HWP Mueller matrix coefficients.

\subsection{Comparison with design parameters}\label{sec:physical_model}

While the major goal of this work is to provide an avenue for including the measured non-idealities of the BLASTPol HWP {\it as-built} in a map-making algorithm, it is useful at this stage to compare the nominal values of the build parameters we assumed to design the HWP with the actual values that can be estimated via the physical and analytical model developed by \citet[][which we refer to for a complete account of the formalism and implementation]{Savini2006,Savini2009}.

\begin{figure}
\centering
\includegraphics[width=1.\linewidth]{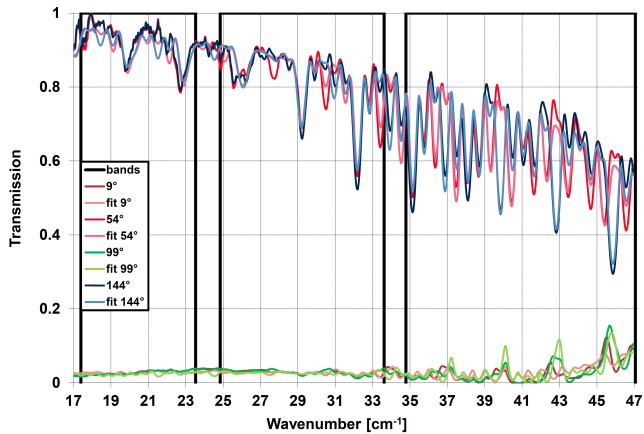}
\caption{Comparison between the measured co-pol transmission spectra near the HWP maxima/minima and the corresponding physical model of \citet{Savini2006}, which is fit to the data by allowing the build parameters of the HWP (refraction index and thickness of all the materials, orientation of the birefringent substrates) to vary in a physical way around the nominal values.}
\label{fig:BLAST_pol_spareHWP_ARC_copol_spectra_Giorgio}
\end{figure}

In this work, the physical model is fit to the spectral data described in Section~\ref{sec:measurements_and_results} by allowing the HWP build parameters to vary in a physical way around the nominal values. We report in Fig.~\ref{fig:BLAST_pol_spareHWP_ARC_copol_spectra_Giorgio} a comparison between the measured co-pol transmission spectra near the HWP maxima/minima and the corresponding physical model, whose best-fit parameters are listed in Table~\ref{tab:physical_parameters_HWP}, along with the nominal ones.

\begin{table}
  \centering
  \begin{tabular}{ccccc}
   \hline
   parameter & units & nominal & best-fit & uncertainty\\
  \hline
 $n^0_o$ & & 3.052\footnotemark[7] & 3.065 & 0.002 \\
 $n^1_o$ & cm & & 0.00207 & 0.00008\\
 $n^0_e$ & & 3.372\footnotemark[7] & 3.444 & 0.001\\
 $n^1_e$ & cm & & 0.00127 & 0.00008\\
$d$ & $\mu$m & 500.0 & 490.4 & 0.5\\
$t_{\rm PE}$ & $\mu$m & 6.0& 3.5 & 0.5\\
$\Phi_1$ & deg & 26& 22 & 1\\
$\Phi_2$ & deg & 90.3& 87 & 1\\
$\Phi_3$ & deg & 26& 23 & 1\\
$\Phi_4$ & deg & 0& -1 & 1\\
$n_{\rm ARC}^1$ &  & 1.375\footnotemark[8] & 1.26 & 0.01\\
$t_{\rm ARC}^1$ & $\mu$m & 54 & 72 & 2\\
$n_{\rm ARC}^2$ & & & 1.62 & 0.01\\
$t_{\rm ARC}^2$ & $\mu$m & & 37 & 3\\
$n_{\rm ARC}^3$ &  & & 2.28 & 0.01\\
$t_{\rm ARC}^3$ & $\mu$m & & 36 & 2\\
 \hline
 \end{tabular}
 \caption{Best-fit build parameters of the BLASTPol HWP estimated using the physical model of \citet{Savini2006}. $n^0_{e(o)}$ and $n^1_{e(o)}$ are, respectively, the constant and linear terms of the refraction index for the sapphire (extra)ordinary axis, modelled with a linear dependence on wavenumber; $d$ is the thickness of the sapphire substrates; $t_{\rm PE}$ is the thickness of the polyethylene layers; $\Phi_i$ ($i=1$, ..., 4) are the four angles at which the sapphire substrates are oriented (assuming $\Phi_0=0^{\circ}$); $n_{\rm ARC}^i$ and $t_{\rm ARC}^i$ ($i=1$, 2, 3) are, respectively, the refraction indeces and the thicknesses of the i-th layer of ARC (in order of penetration into the HWP), which is modelled as series of equivalent dielectrics.}\footnotetext{7}{$^7$\citet{Loewenstein1973} ~~~$^8$PTFE \citet{Zhang2009}}
 \label{tab:physical_parameters_HWP}
\end{table}
\setcounter{footnote}{8}

Among all the parameters presented in Table~\ref{tab:physical_parameters_HWP}, let us focus on the angles at which the sapphire substrates are oriented, $\Phi_i$ ($i=1$, ..., 4; see Section~\ref{sec:HWP_manu}), since they have a greater impact on what follows. The best-fit orientation angles are up to 3--4$^\circ$ smaller than the design goal. While this is not surprising given the practical challenge of keeping a stack of five plates (interspersed with thin slippery layers of polyethylene) aligned to within $\sim$3\,mm (linear length of a 3.5$^\circ$ arc for a 100\,mm diameter), such a deviation from the desired values will affect the HWP performance, and in particular its phase shift as a function of frequency.

In the following sections we present all the HWP performance parameters
, for which the physical model presented in this section is verified to be in general agreement with the empirical model we develop in Section~\ref{sec:empirical_model}. However, we will show in Section~\ref{sec:phase_shift} that the empirical model is inadequate to retrieve the HWP phase shift and we will have to resort to the physical model again to compare the design and best-fit phase shift versus frequency. This will give us a chance to expand more on which performance parameters are more directly affected by plate alignment errors.

\subsection{Modulation function and efficiency}\label{sec:modulation_function}

\begin{figure}
\centering
\includegraphics[width=1.\linewidth]{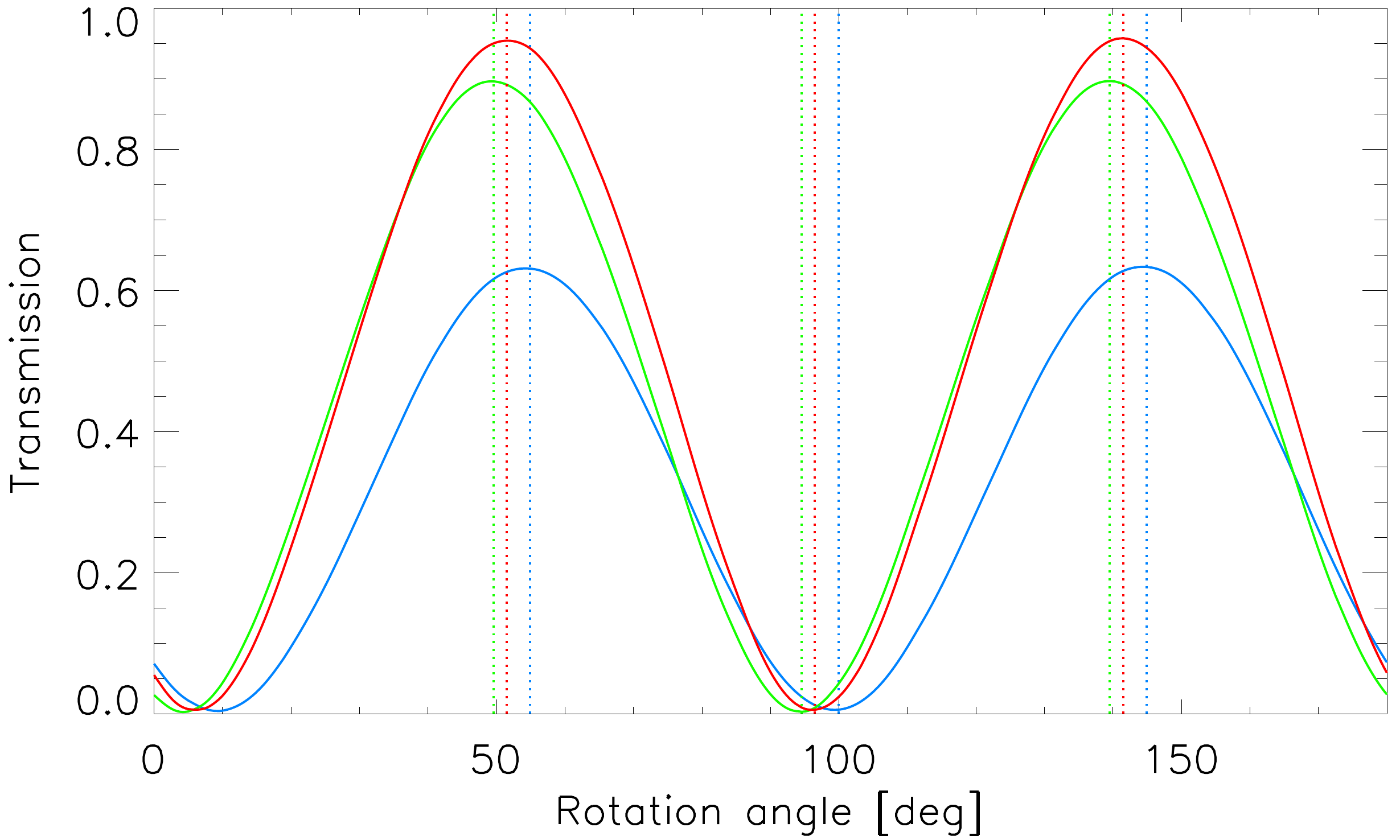}
\caption{Band-integrated co-pol modulation functions of the
BLASTPol HWP at $\sim$120\,K. The curves show the HWP polarisation
modulation functions for a fully polarised source (with a flat
spectrum) parallel to the analyser in the three spectral bands.
Note how the positions of the maxima (and minima) depend on the
wavelength, even when considering a flat-spectrum polarised input
source; the dotted vertical lines show the band-integrated
positions of the HWP extrema (shown in
Fig.~\ref{fig:HWP_offset_vs_freq}), which result from the fitting
routine described in the next sections.}
\label{fig:modulation_curves_copol_spareHWP}
\end{figure}

\begin{figure}
\centering
\includegraphics[width=1.\linewidth]{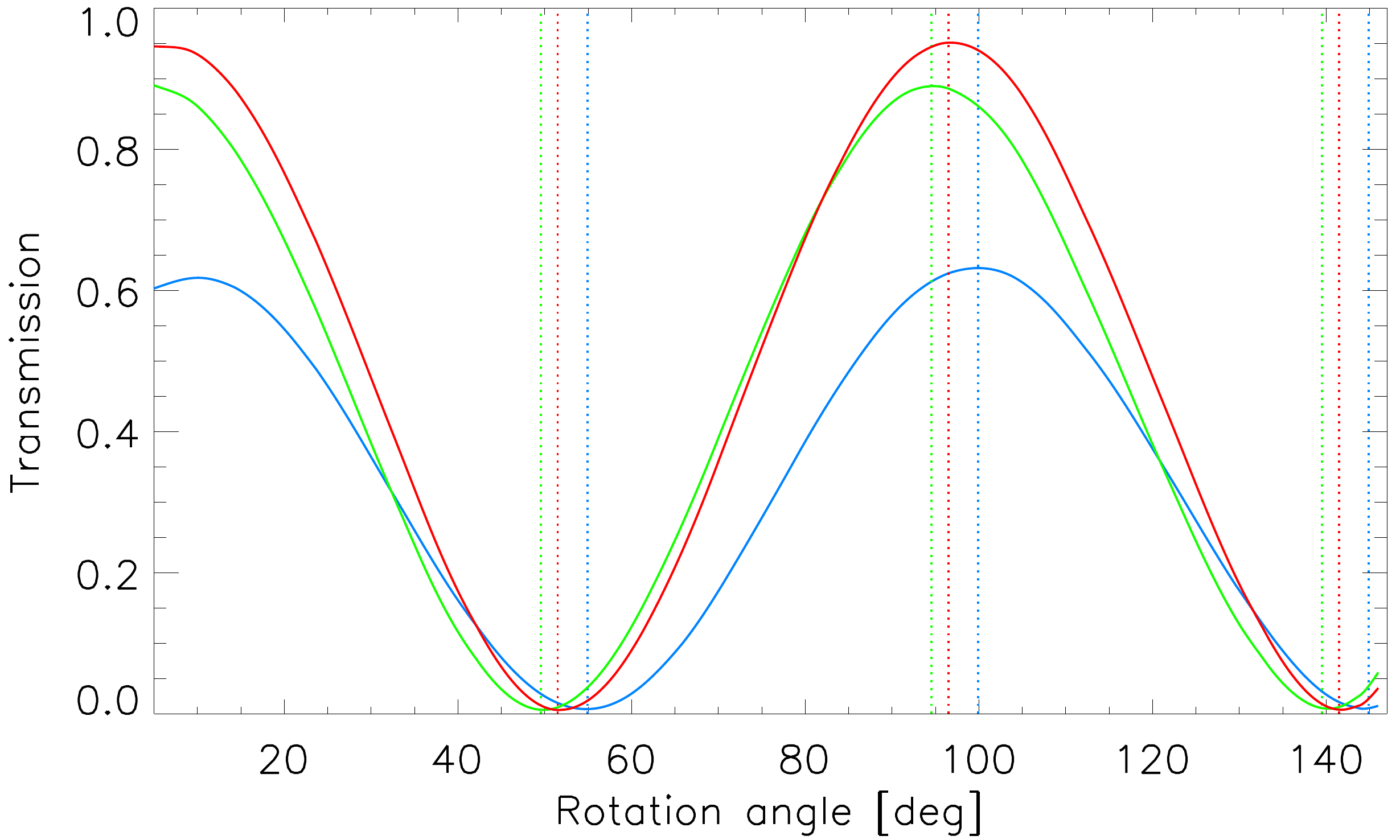}
\caption{Band-integrated modulation functions equivalent to those
shown in Fig.~\ref{fig:modulation_curves_copol_spareHWP} but for
cross-pol transmission.}
\label{fig:modulation_curves_xpol_spareHWP}
\end{figure}

We can reduce the dependence on frequency of our data cubes by
integrating over the spectral bands of BLASTPol, as follows:
\begin{equation}\label{eq:modulation_curve}
\overline{T}^{\rm ch}_{\rm cp} \left(\theta\right)=
\frac{\int_0^{\infty} \Sigma^{\rm ch}\left(\nu\right)\,T_{\rm
cp}\left(\theta,\nu\right)\,d\nu}{\int_0^{\infty} \Sigma^{\rm
ch}\left(\nu\right)\,d\nu}~.
\end{equation}
Here the superscript ``ch'' refers to one among 250, 350, and
500\,$\mu$m, $\Sigma^{\rm ch}\left(\nu\right)$ is the measured
spectral response of each of the BLASTPol bands
\citep[see][]{Pascale2008}, and $T_{\rm
cp}\left(\theta,\nu\right)$ are points on the co-pol surface
depicted in Fig.~\ref{fig:spareHWP_copol_spectra_surface}. A
similar expression can be written for the cross-pol
band-integrated transmission. By performing this integration at
every angle for which spectral data have been obtained, the
interpolation of these data points will result in the modulation
functions of the HWP at $\sim$120\,K for each of the BLASTPol bands; these curves are shown in
Figs.~\ref{fig:modulation_curves_copol_spareHWP} (co-pol) and~\ref{fig:modulation_curves_xpol_spareHWP} (cross-pol).

The modulation curves presented here are valid for input sources that have a flat spectrum in the BLASTPol bands. Equation~\ref{eq:modulation_curve} can be generalised to include
the known (or assumed) spectral signature of a given astronomical
or calibration source \citep[see also
e.g.,][Equation~2]{Novak1989a}. More generally, all the
band-averaged quantities that we have defined here and will be
defined in the following are potentially affected by the spectral
shape of the input source. However, we will see how the HWP
transmission and modulation efficiency are very weakly dependent
on the spectral index of the input source, whereas the position of
the equivalent axes of the sapphire plate stack is more
significantly affected \citep[see also the analysis carried out
by][]{Savini2009}, especially at 250 and 500\,$\mu$m.

Figs.~\ref{fig:modulation_curves_copol_spareHWP} and
\ref{fig:modulation_curves_xpol_spareHWP} clearly show that there
is a significant dependence of the position of the HWP maxima and
minima upon frequency, even when considering a flat-spectrum
polarised input source. These effects are particularly important
for a ``HWP step and integrate'' experiment such as BLASTPol, and a polarisation calibration must be performed by using information from the characterisation of the HWP. We begin to tackle this problem in the next section, where we outline a relatively simple solution to account for most of the HWP non-idealities in the data-analysis pipeline, and in particular in the map-making algorithm (see
Section~\ref{sec:map_maker}).

The spectral transmission datasets of the HWP cooled to
$\sim$120\,K, when compared to those taken with the HWP at room
temperature (not reported here for brevity), show a
definite abatement of the in-band losses due to absorption from
sapphire, as expected. However the effect is still appreciable,
especially above $\sim$25\,cm$^{-1}$. As we will show in the
following, we have evidence that the residual
absorption nearly vanishes when the sapphire is further cooled to
4\,K, as it is when the HWP is installed in the BLASTPol cryostat.
While it is not currently feasible for us to measure the spectral
response of the HWP cooled to 4\,K, the unique quality and
completeness of our dataset allow us to fully characterise the
performance of the BLASTPol HWP.

\begin{figure}
\centering
\includegraphics[width=1.\linewidth]{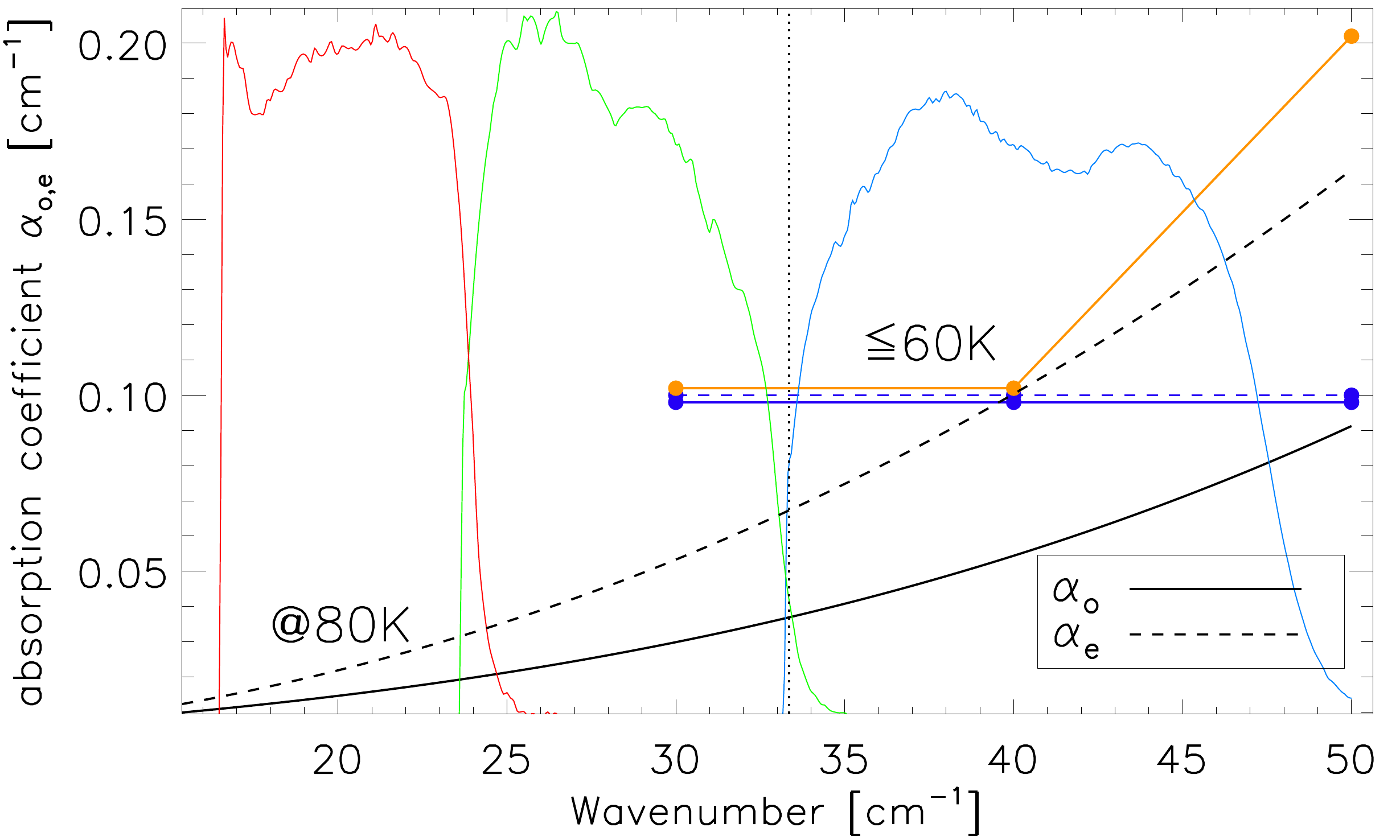}
\caption{Ordinary (solid) and extraordinary (dashed)
sapphire absorption coefficient as a function of wavenumber, at
cryogenic temperatures. The two analytical relations covering the
whole frequency range are derived by Savini (2010, pers. comm.) from a set of spectral
measurements of a sample at 80\,K, and, strictly
speaking, only apply at frequencies $\lesssim$\,1\,THz (dotted vertical line). For reference, we plot measurements from
\citet[][purple]{Loewenstein1973} at 1.5\,K and
\citet[][ochre]{Cook1985} at 60\,K, displaced by 0.002\,cm$^{-1}$ in $y$ for visual clarity; the lines connecting these data points follow the convention shown
in the legend.}\label{fig:sapphire_absorption_coefficient_cold}
\end{figure}

\begin{figure}
\centering
  \subfigure[The derived transmissions through the cold HWP as a
function of wavenumber. The black line shows the HWP transmission,
$T^{0^{\circ}}_{\rm cp}$, between two parallel polarisers ($Q=1
\rightarrow Q=1$) with the HWP near one maximum (0$^{\circ}$). The dark blue
line shows $Q=-1 \rightarrow Q=-1$ in the same reference frame (or
equivalently $Q=1 \rightarrow Q=1$ with the HWP axis at
90$^{\circ}$). The purple line shows the transmission,
$T^{0^{\circ}}_{\rm xp}$, with the HWP axis at 0$^{\circ}$ between
two perpendicular polarisers.]
  {\includegraphics[width=0.92\linewidth, keepaspectratio]{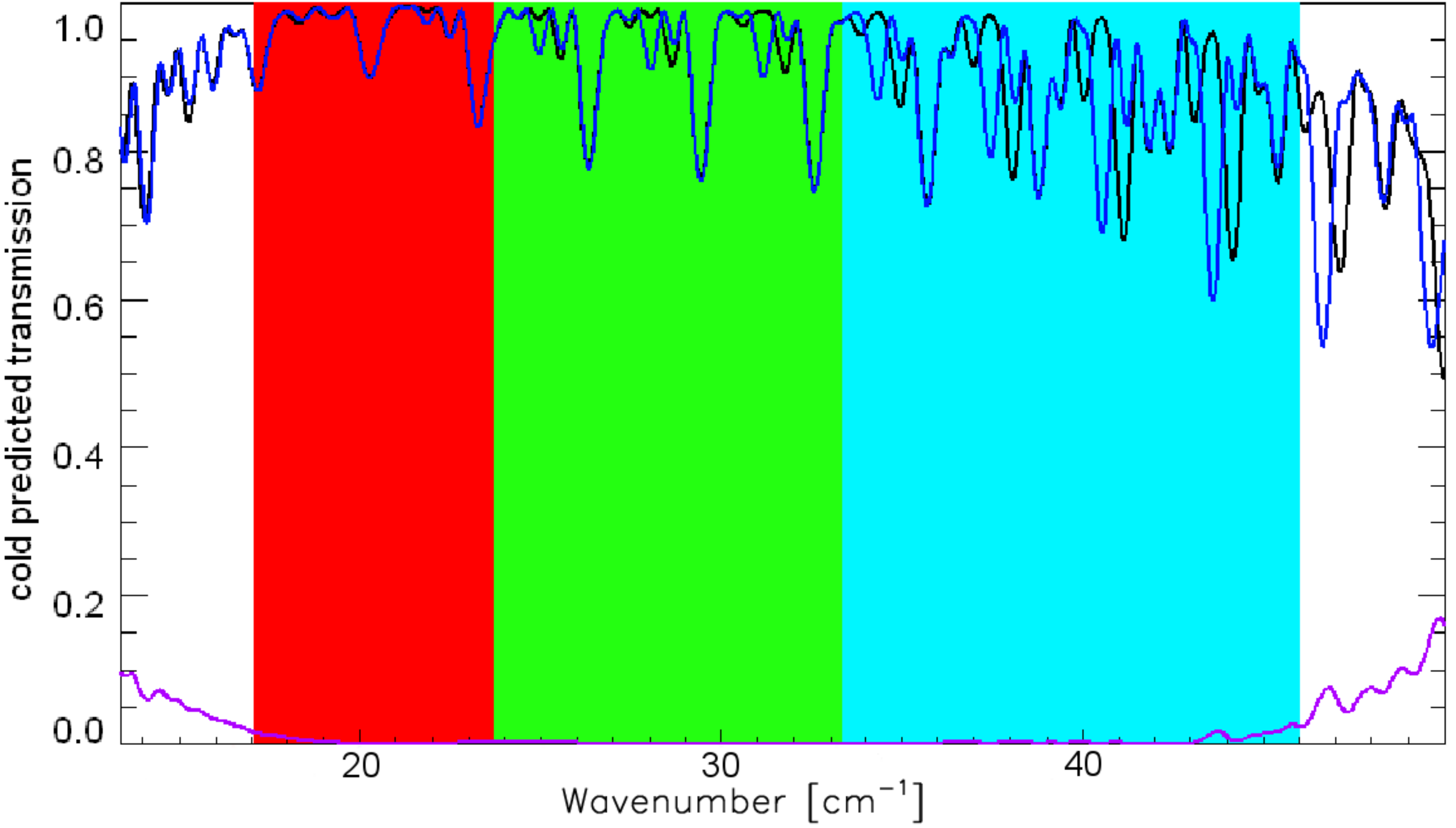}}
  \subfigure[Derived modulation efficiency of the cold HWP as a
function of wavenumber, obtained as ($T^{0^{\circ}}_{\rm
cp}-T^{0^{\circ}}_{\rm xp}$)/($T^{0^{\circ}}_{\rm
cp}+T^{0^{\circ}}_{\rm xp}$). Note that the $y$-axis scale ranges
from 0.8 to 1.]
  {\includegraphics[width=0.92\linewidth, keepaspectratio]{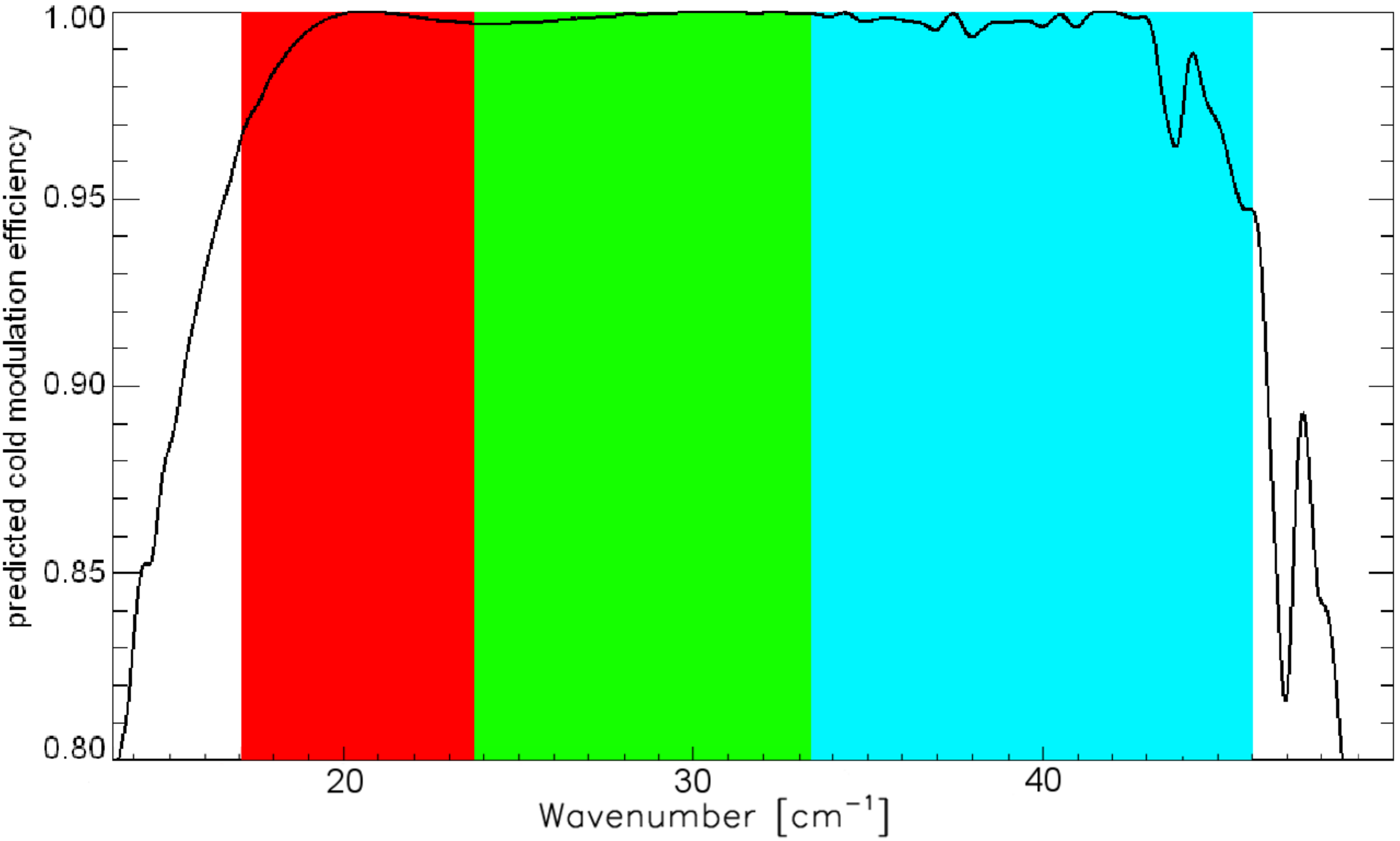}}
\caption{Derived performance of the BLASTPol HWP at 4\,K,
extrapolated from a set of spectral data collected with the HWP
cooled to $\sim$120\,K (see Section~\ref{sec:measurements_and_results}).
``Co-pol'' and ``cross-pol'' transmissions, $T_{\rm cp}$ and
$T_{\rm xp}$, are defined as in
Fig.~\ref{fig:FTS_scheme_HWP_measurements}. The approximate extent
of the BLASTPol bands is also indicated.}\label{fig:modeff}
\end{figure}

We extrapolate our ``cold'' dataset to 4\,K, using the analytical relations shown in Fig.~\ref{fig:sapphire_absorption_coefficient_cold}\footnote{The analytical relations apply, strictly speaking, at 80\,K and for
$k\lesssim33$\,cm$^{-1}$, thus we corroborate them at higher
frequencies with the data points, which apply at $\lesssim$60\,K. It is
evident that the sapphire absorption coefficient has a very weak
dependence on temperature below 80\,K, and in particular data points
collected at 1.5\,K are in good enough agreement (within 2\% on
the resulting absorption for $d=2.5$\,mm) with those collected at
higher temperatures (up to 80\,K). Therefore we can safely claim
that for our application the two analytical relations shown in
Figure~\ref{fig:sapphire_absorption_coefficient_cold} are a good
representation of the sapphire absorption at 4\,K.}. The HWP
modulation efficiency is defined as ($T^{0^{\circ}}_{\rm
cp}-T^{0^{\circ}}_{\rm xp}$)/($T^{0^{\circ}}_{\rm
cp}+T^{0^{\circ}}_{\rm xp}$), where the ``co-pol'' and
``cross-pol'' transmissions, $T^{0^{\circ}}_{\rm cp}$ and
$T^{0^{\circ}}_{\rm xp}$, are the spectral responses
of the HWP near one of the transmission maxima (called 0$^{\circ}$ here), between parallel and
perpendicular polarisers, respectively. The inferred
co-pol/cross-pol transmissions and modulation efficiency of the
BLASTPol HWP (with its axis at 0$^{\circ}$) at 4\,K are shown in
Fig.~\ref{fig:modeff}. For a flat-spectrum input source, the
band-integrated transmission of the HWP at its maxima is
$\sim$0.87, $\sim$0.91, and $\sim$0.95 at 250, 350, and
500\,$\mu$m, respectively; whereas the band-integrated cross-pol
is $\la$\,0.5\%, $\la$\,0.2\%, and $\la$\,0.5\%, respectively;
finally, the band-integrated modulation efficiency is $\sim$98.8\%
$\sim$99.5\%, and $\sim$99.0\%, respectively.

\section{Empirical modelling}\label{sec:empirical_model}

The final goal of this section is to provide a set of usable
parameters that completely describe the performance of the HWP as
measured in the laboratory. This set of parameters consists of the
16 coefficients of the Mueller matrix of a generic HWP, and the
actual phase shift. For an ideal HWP, the Mueller matrix at
$\theta=0^{\circ}$ reads \citep{Collett1993}
\begin{equation}\label{eq:ideal_HWP}
{\mathbfss M_{\rm HWP}}=
\begin{pmatrix}
  1 & 0 & 0 & 0 \\
  0 & 1 & 0 & 0 \\
  0 & 0 & -1 & 0 \\
  0 & 0 & 0 & -1 \\
\end{pmatrix}~,
\end{equation}
and the phase shift is $\Delta\varphi=180^{\circ}$.

For a real HWP, these parameters always depart from the ideal case
to some extent, and certainly depend upon frequency. In the
following we describe an empirical model that we develop
specifically for the characterisation of the BLASTPol HWP, though
we note that it can be applied to any HWP to recover its
frequency-dependent descriptive parameters. Such an empirical
model is complementary to the physical and analytical one
developed by \citet{Savini2006,Savini2009}, which produces an
analogous output by modelling the non-idealities of the components
of the HWP assembly and their optical parameters.

\subsection{Mueller matrix characterisation}\label{sec:mueller_HWP}

By recalling the Stokes formalism, we can formalise the
experimental apparatus described in
Section~\ref{sec:cold_spectra} as a series of matrix products, as
follows:
\begin{eqnarray}
S^{\rm\,cp}_{\rm out}&=&\vec{\mathbfit D}^{\rm T}\,\cdot\,{\mathbfss M^{\rm h}_{\rm p}}\,\cdot\,{\mathbfss R}\left(-\theta\right)\,\cdot\,{\mathbfss M_{\rm HWP}}\,\cdot\,{\mathbfss R}\left(\theta\right)\,\cdot\vec{\mathbfit S}^{\rm\,h}_{\rm in}~;\label{eq:mueller_HWP_1a}\\
S^{\rm\,xp}_{\rm out}&=&\vec{\mathbfit D}^{\rm
T}\,\cdot\,{\mathbfss M^{\rm v}_{\rm p}}\,\cdot\,{\mathbfss
R}\left(-\theta\right)\,\cdot\,{\mathbfss M_{\rm
HWP}}\,\cdot\,{\mathbfss
R}\left(\theta\right)\,\cdot\vec{\mathbfit S}^{\rm\,h}_{\rm
in}~.\label{eq:mueller_HWP_1b}
\end{eqnarray}
Here $\vec{\mathbfit D}$ is the Stokes vector for a bolometric
(polarisation insensitive) intensity detector, ${\mathbfss M^{\rm
h}_{\rm p}}$ is the Mueller matrix of an ideal horizontal
polariser, ${\mathbfss M^{\rm v}_{\rm p}}$ is that of an ideal
vertical polariser, ${\mathbfss R}\left(\theta\right)$ is the
generic Mueller rotation matrix, and $\vec{\mathbfit S}_{\rm in}$
is the horizontally polarised input beam from the pFTS. By
expanding all the matrices in Equation~\ref{eq:mueller_HWP_1a},
\begin{eqnarray}\label{eq:mueller_HWP_2}
S^{\rm\,cp}_{\rm out}=\begin{pmatrix}
 1 & 0 & 0 & 0
\end{pmatrix}\cdot
\begin{pmatrix}
  1 & 1 & 0 & 0 \\
  1 & 1 & 0 & 0 \\
  0 & 0 & 0 & 0 \\
  0 & 0 & 0 & 0 \\
\end{pmatrix}\cdot~~~~~~~~~~~~~~~~\\
\nonumber\cdot
\begin{pmatrix}
  1 & 0 & 0 & 0 \\
  0 & \cos(2\theta) & \sin(2\theta) & 0 \\
  0 & -\sin(2\theta) & \cos(2\theta) & 0 \\
  0 & 0 & 0 & 1 \\
\end{pmatrix}
\cdot
\begin{pmatrix}
  a_{00} & a_{01} & a_{02} & a_{03} \\
  a_{10} & a_{11} & a_{12} & a_{13} \\
  a_{20} & a_{21} & a_{22} & a_{23} \\
  a_{30} & a_{31} & a_{32} & a_{33} \\
\end{pmatrix}\cdot\\
\nonumber\cdot
\begin{pmatrix}
  1 & 0 & 0 & 0 \\
  0 & \cos(2\theta) & -\sin(2\theta) & 0 \\
  0 & \sin(2\theta) & \cos(2\theta) & 0 \\
  0 & 0 & 0 & 1 \\
\end{pmatrix}
\cdot
\begin{pmatrix}
  1 \\
  1 \\
  0 \\
  0 \\
\end{pmatrix}~,~~~~~~~~~~~~~~~~~~~~~~~~
\end{eqnarray}
we can compute the products and rearrange as
\begin{eqnarray}
S^{\rm\,cp}_{\rm out}=\frac{1}{2} \Big[2 a_{00}+a_{11}+a_{22}+2
(a_{01}+a_{10})
\cos2\theta+\label{eq:mueller_HWP_4a}~~~~~\\
\nonumber+(a_{11}-a_{22}) \cos4\theta-2 (a_{02}+a_{20})
\sin2\theta -(a_{12}+a_{21}) \sin4\theta\Big]\\
=A+B \sin2\theta +C \cos2\theta + D \sin4\theta + E
\cos4\theta~,~~~~~\label{eq:mueller_HWP_4b}
\end{eqnarray}
with
\begin{eqnarray}\label{eq:mueller_HWP_5}
\nonumber A &\equiv& a_{00}+\frac{a_{11}}{2}+\frac{a_{22}}{2}~,\\
B&\equiv&-\left(a_{02} + a_{20}\right),~~~~C\equiv a_{01}+a_{10}~,\\
\nonumber D&\equiv&-\frac{1}{2}\left(a_{12} +
a_{21}\right),~~~~E\equiv\frac{1}{2}\left(a_{11}-a_{22}\right)~.
\end{eqnarray}
Similarly, noting that
\begin{equation*}
{\mathbfss M^{\rm v}_{\rm p}}=\begin{pmatrix}
  1 & -1 & 0 & 0 \\
  -1 & 1 & 0 & 0 \\
  0 & 0 & 0 & 0 \\
  0 & 0 & 0 & 0 \\
\end{pmatrix}~,
\end{equation*}
we rearrange Equation~\ref{eq:mueller_HWP_1b} as
\begin{eqnarray}
S^{\rm\,xp}_{\rm out}=A'+B' \sin2\theta +C' \cos2\theta + D'
\sin4\theta + E' \cos4\theta~,~~~~~\label{eq:mueller_HWP_8b}
\end{eqnarray}
with
\begin{eqnarray}\label{eq:mueller_HWP_9}
\nonumber A' &\equiv& a_{00}-\frac{a_{11}}{2}-\frac{a_{22}}{2}~,\\
B'&\equiv& a_{20} - a_{02},~~~~C'\equiv a_{01}-a_{10}~,\\
\nonumber D'&\equiv&\frac{1}{2}\left(a_{12} +
a_{21}\right),~~~~E'\equiv\frac{1}{2}\left(a_{22}-a_{11}\right).
\end{eqnarray}

Finally, by performing linear combinations of the quantities
defined in Equations~\ref{eq:mueller_HWP_5} and
\ref{eq:mueller_HWP_9}, one can write the individual elements that
compose the Mueller matrix of a generic HWP as
\begin{eqnarray}\label{eq:mueller_HWP_10}
\nonumber a_{00} &=& \frac{1}{2}\,\left(A+A'\right),~~~a_{01}=\frac{1}{2}\,\left(C+C'\right)~,\\
a_{10} &=& \frac{1}{2}\,\left(C-C'\right),~~~a_{11}=\frac{1}{2}\,\left(A-A'+E-E'\right)~,\\
\nonumber a_{02} &=& -\frac{1}{2}\,\left(B+B'\right),~~~a_{20}=\frac{1}{2}\,\left(B'-B\right)~,\\
\nonumber a_{22} &=&
\frac{1}{2}\,\left(A-A'-E+E'\right),~~~a_{12}=a_{21}=\frac{1}{2}\,\left(D'-D\right)~,
\end{eqnarray}
where in the last equality we currently assume the symmetry of two
coefficients, $a_{12}=a_{21}$. This degeneracy may be broken by
imposing the conservation of energy, i.e. by requiring the output
Stokes vector resulting from a generic polarised input travelling
through the recovered HWP Mueller matrix to satisfy $I^2 = Q^2 +
U^2$. Alternatively, the degeneracy can be broken by taking
spectra at an intermediate configuration between co- and
cross-pol; this additional constraint will be included in a future
work (Spencer et al., in preparation). Also, because our
experimental setup is sensitive to linear but not circular
polarisation, this method only allows us to constrain the 9
elements of the Mueller matrix associated with
$\left[I,Q,U\right]$. The remaining 7 coefficients associated with
$V$ can only be measured with the use of a quarter-wave plate,
which induces a phase shift of 90$^{\circ}$ between the two
orthogonal polarisations travelling through the plate; this
measurement is beyond the scope of this paper and not pertinent to
the needs of BLASTPol.

We want to estimate the 9 coefficients derived in
Equation~\ref{eq:mueller_HWP_10} from the co-pol and cross-pol
data cubes described in Section~\ref{sec:measurements_and_results}.
Equations~\ref{eq:mueller_HWP_4b} and \ref{eq:mueller_HWP_8b}
encode a simple dependence of $S^{\rm\,cp}_{\rm out}$ and
$S^{\rm\,xp}_{\rm out}$ upon $\theta$, the HWP rotation angle.
Therefore, for a given frequency, a fitting routine can be applied
to the measured transmission curves as a function of $\theta$, to
determine the parameter sets $[A,B,C,D,E]$ and $[A',B',C',D',E']$
for the co-pol and cross-pol configurations, respectively. By
repeating the fit for every frequency, we have an estimate of the
9 coefficients as a function of wavelength. However, this
procedure does not allow us to associate an uncertainty to our
estimates.

A better approach to this problem is to use a Monte Carlo
simulation. We repeat the above fitting procedure 1000 times;
every time we add to every individual transmission curve a
realisation of white noise, scaled to the 1\,$\sigma$ spectral
uncertainty as estimated in Fig.~\ref{fig:plot_noise_backs_HWP},
and compute the fit using this newly generated transmission curve.
In addition, for every frequency we introduce a random jitter on
the rotation angle that has a 1\,$\sigma$ amplitude of
1$^{\circ}$. The dispersion in the fitted parameters due to these
two types of uncertainties, which are inherent to the measurement
process, provides a realistic estimate of the uncertainty
associated with each of the 9 coefficients. In particular, at each
frequency, we produce 9 histograms of the 1000 fitted values. We
use the mode of each distribution as our best estimate for the
corresponding coefficient at that frequency, and the 68\%
confidence interval as the associated 1\,$\sigma$ error.

\begin{figure}
\centering
\includegraphics[width=1.\linewidth]{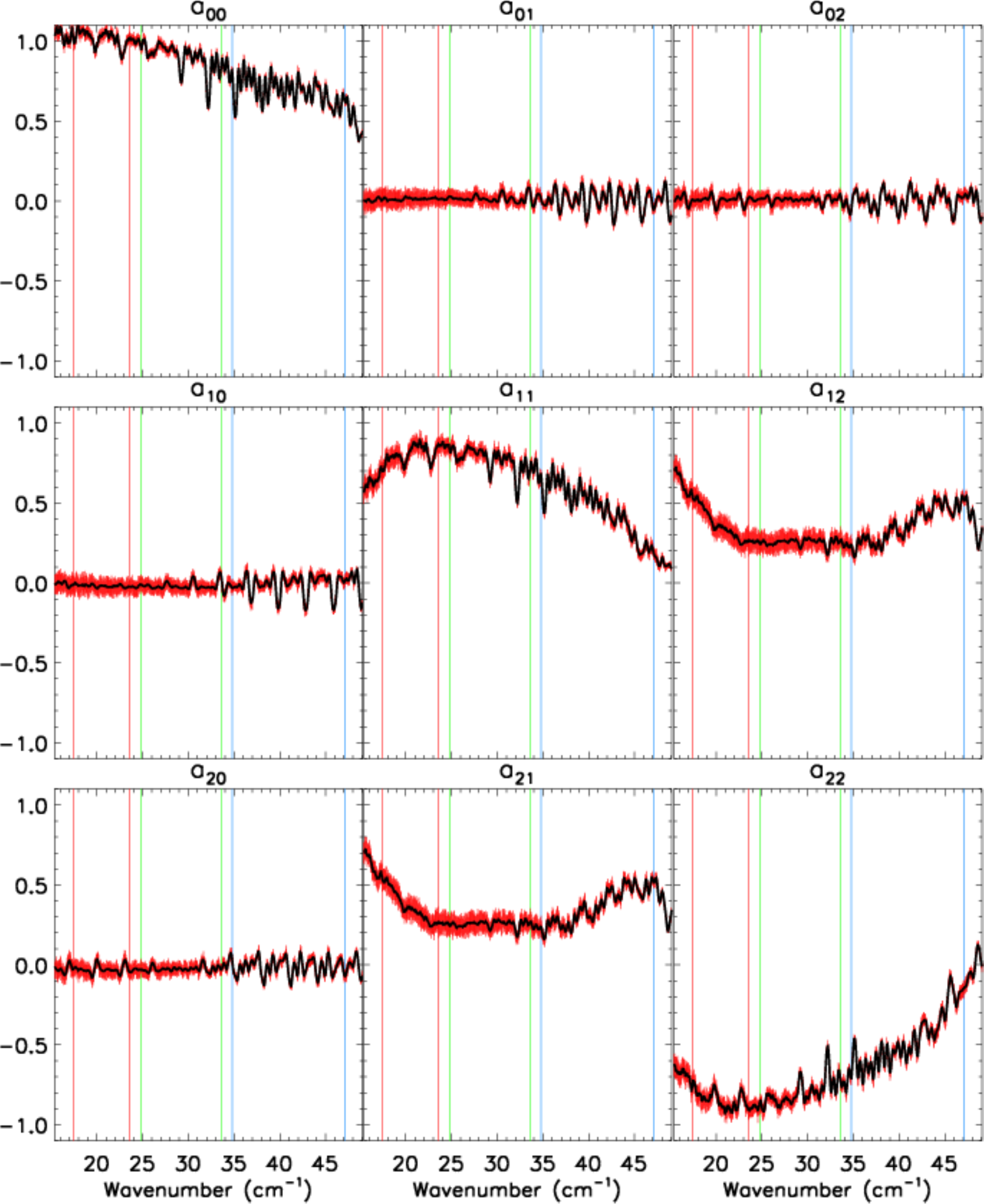}
\caption{Graphical representation of the Mueller matrix of the
BLASTPol HWP at a given angle ($\theta=0^{\circ}$), as a function
of wavenumber. The (10\,$\sigma$) error bars (in red) are
quantified via a Monte Carlo, which accounts for random errors in
the spectra of amplitude as given in
Fig.~\ref{fig:plot_noise_backs_HWP}, and random errors of
amplitude 1$^{\circ}$ in the rotation angle.}
\label{fig:spare_HWP_MM_coeff_no_offset}
\end{figure}

\begin{figure}
\centering
\includegraphics[width=1.\linewidth]{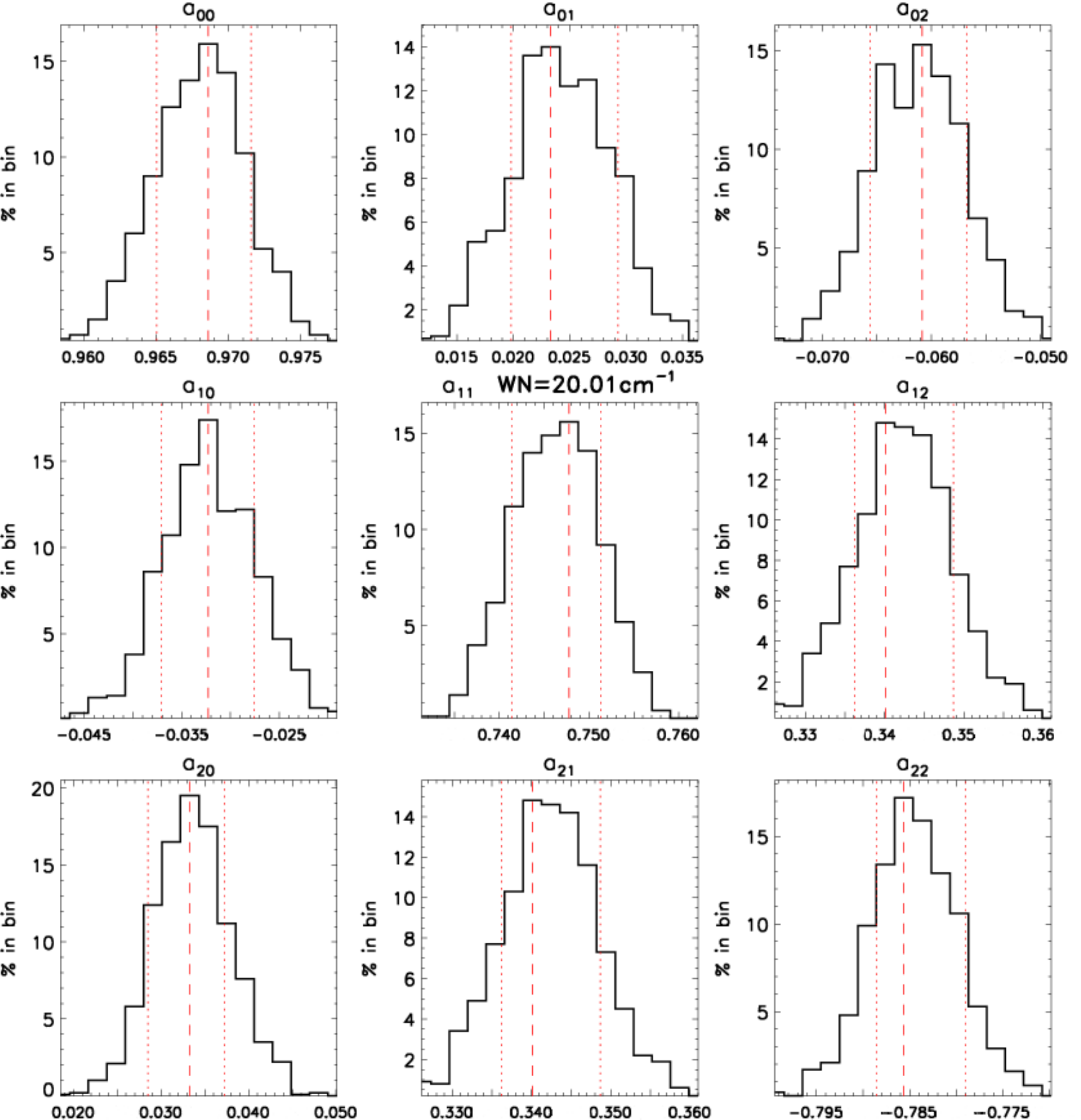}
\caption{Histograms at 20\,cm$^{-1}$ (central frequency of the
500\,$\mu$m BLASTPol band) resulting from the Monte Carlo fit of
the HWP parameters. For every histogram, the dashed red line
indicates the mode of the distribution, which we adopt as our best
estimate for the corresponding coefficient at that frequency,
while the two dotted red lines indicate the 68\% confidence
interval, which we use as the uncertainty on the retrieved
coefficient.} \label{fig:histo_spare_HWP_MM_histos_20.01cm-1}
\end{figure}

In Fig.~\ref{fig:spare_HWP_MM_coeff_no_offset} we show a graphical
representation of the 9-element Mueller matrix of the BLASTPol HWP
at a given angle ($\theta=0^{\circ}$), as a function of
wavenumber. In
Fig.~\ref{fig:histo_spare_HWP_MM_histos_20.01cm-1} we show the
resulting histograms for the 9 coefficients at 20\,cm$^{-1}$, central frequency of the 500\,$\mu$m BLASTPol band; histograms at 28.57\,cm$^{-1}$ (350\,$\mu$m) and 40.02\,cm$^{-1}$ (250\,$\mu$m) look very similar and thus are not presented here for brevity.

\subsection{Position of the HWP equivalent axes}\label{sec:beta_ea}

\begin{figure}
\centering
\includegraphics[width=1.\linewidth]{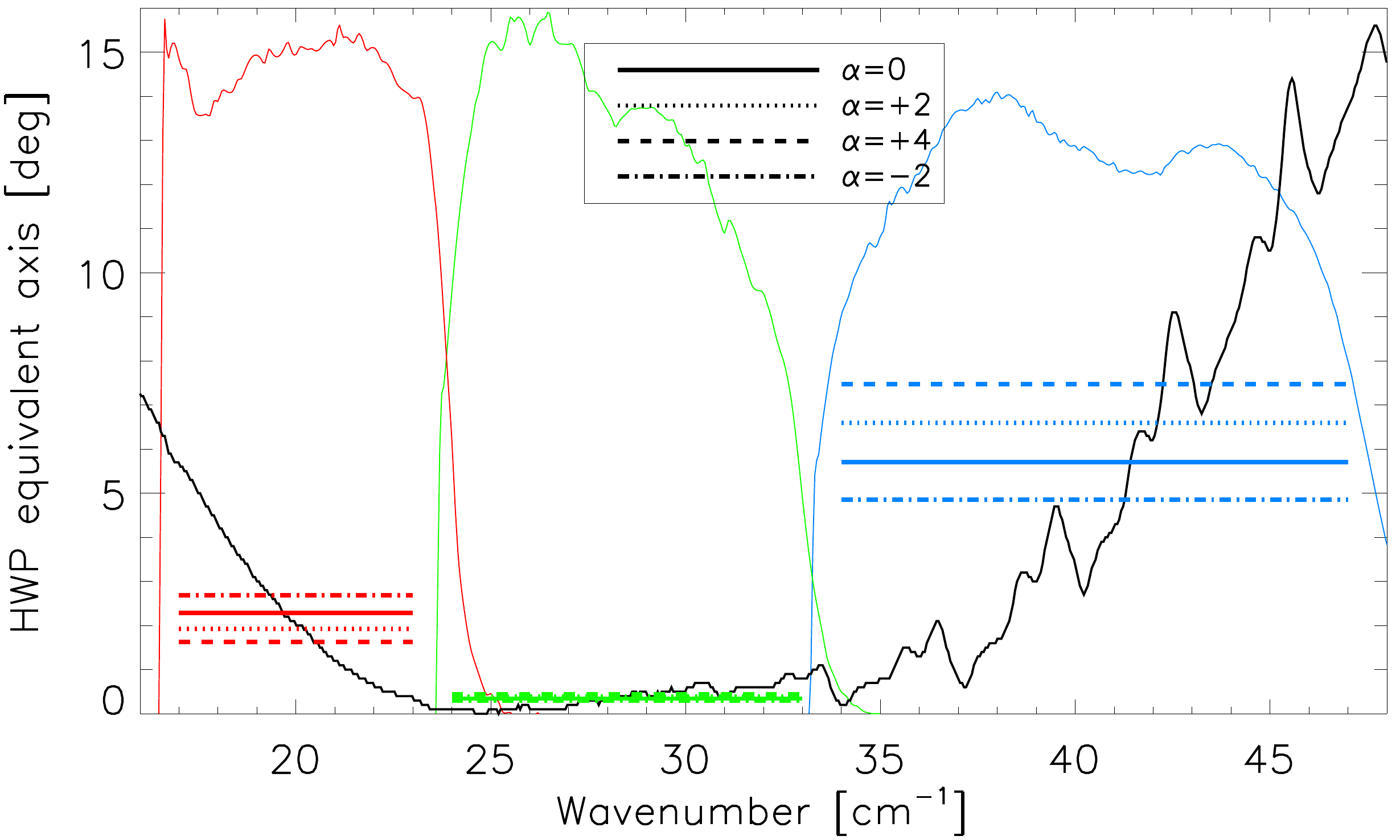}
\caption{Position of the HWP equivalent axis, $\beta_{\rm ea}$, as
a function of wavenumber (solid black line). Note that this
quantity is defined with respect to an arbitrary constant offset
that is inherent to the specific experimental setup; we set this
offset to be zero at 25\,cm$^{-1}$. The band-averaged values for
input sources with different spectral index ($\alpha$; see legend)
are drawn as thick horizontal lines. Also shown for reference is
the relative spectral response of the three BLASTPol channels, in
arbitrary units.} \label{fig:HWP_offset_vs_freq}
\end{figure}

The behaviour of the coefficients as a function of wavenumber
shown in Fig.~\ref{fig:spare_HWP_MM_coeff_no_offset} confirms that
the position of the HWP equivalent axes, $\beta_{\rm ea}$
hereafter, has an inherent frequency dependence, which we
must investigate. $\beta_{\rm ea}$ can be readily retrieved at
each frequency by locating the rotation angle that corresponds to
the first minimum in the fitted transmission curve. Hence,
$\beta_{\rm ea}$ is measured with respect to an arbitrary constant
offset that is inherent to the specific experimental setup; we set
this offset to be zero at 25\,cm$^{-1}$. Operationally, this means
that the HWP zero angle in the instrument reference frame
($\beta_0$; see Equation~\ref{eq:polangle}) must be calibrated
using the 350\,$\mu$m band. A plot of $\beta_{\rm ea}$ as a
function of wavenumber is given in
Fig.~\ref{fig:HWP_offset_vs_freq}.

As anticipated, it is of crucial importance to derive the
band-averaged value of $\beta_{\rm ea}$ for input sources with
different spectral signature, as follows:
\begin{equation}\label{eq:beta_ea}
\overline{\beta}_{\rm ea}^{\rm ch}= \frac{\int_0^{\infty}
\Sigma^{\rm ch}\left(\nu\right)\,\beta_{\rm
ea}\left(\nu\right)\,\varsigma\left(\nu\right)\,d\nu}{\int_0^{\infty}
\Sigma^{\rm
ch}\left(\nu\right)\,\varsigma\left(\nu\right)\,d\nu}~,
\end{equation}
where we adopt the same notation as in
Equation~\ref{eq:modulation_curve} and the known (or assumed)
spectrum of an astronomical or calibration source is modelled as
$\varsigma\left(\nu\right) \propto \nu^{\alpha}$. We compute
Equation~\ref{eq:beta_ea} for a range of spectral indices of
interest: $\alpha=0$ for a flat spectrum; $\alpha=2$ for the
Rayleigh-Jeans tail of a blackbody; $\alpha=4$ for interstellar
dust, modelled as a modified blackbody with emissivity $\beta=2$
\citep{Hildebrand1983}; and finally $\alpha=-2$ as a replacement
for the mid-infrared exponential on the Wien side of a blackbody
to account for the variability of dust temperatures within a
galaxy \citep[][]{Blain1999c,Blain2003}. The results of this
analysis are shown in Fig.~\ref{fig:HWP_offset_vs_freq} and in
Table~\ref{tab:HWP_offset_vs_freq}.

\begin{table}
  \centering
  \begin{tabular}{cccc}
   \hline
   &  & $\overline{\beta}_{\rm ea}$ [deg] & \\
   \hline
   $\alpha$ & 250\,$\mu$m & 350\,$\mu$m & 500\,$\mu$m\\
  \hline
 $-$2 & 4.9 & 0.30 & 2.7\\
0  & 5.7 & 0.35 & 2.3\\
$+$2 & 6.6 & 0.39 & 1.9\\
$+$4 & 7.5 & 0.44 & 1.6\\
 \hline
 \end{tabular}
 \caption{Band-averaged position of
the HWP equivalent axis for sources with different spectral index.
The input source is assumed to have a spectrum $\varsigma\propto
\nu^{\alpha}$.}
 \label{tab:HWP_offset_vs_freq}
\end{table}

As expected, the impact of different input spectral signatures is
minimal at 350\,$\mu$m, where the HWP has been designed to
function optimally (see Section~\ref{sec:HWP_manu}); whereas the
spectral dependence is more pronounced at 250 and 500\,$\mu$m,
and, if neglected, it may lead to an arbitrary rotation of the
retrieved polarisation angle on the sky of magnitude
$2\,\overline{\beta}_{\rm ea} =10$--15$^{\circ}$ (3--5$^{\circ}$)
at 250 (500)\,$\mu$m (see Equation~\ref{eq:polangle}).

We have thus confirmed that the dependence of the HWP equivalent
axes upon wavelength is inherent to the achromatic design. We now
postulate that most of the non-idealities we see in the measured
HWP Mueller matrix (Fig.~\ref{fig:spare_HWP_MM_coeff_no_offset})
are primarily due to the wavelength dependence of $\beta_{\rm
ea}$, along with the residual absorption from sapphire at
$\sim$120\,K. 
One can imagine that the HWP
performance would approach the ideal case once this effect is
corrected for.

\begin{figure}
\centering
\includegraphics[width=1.\linewidth]{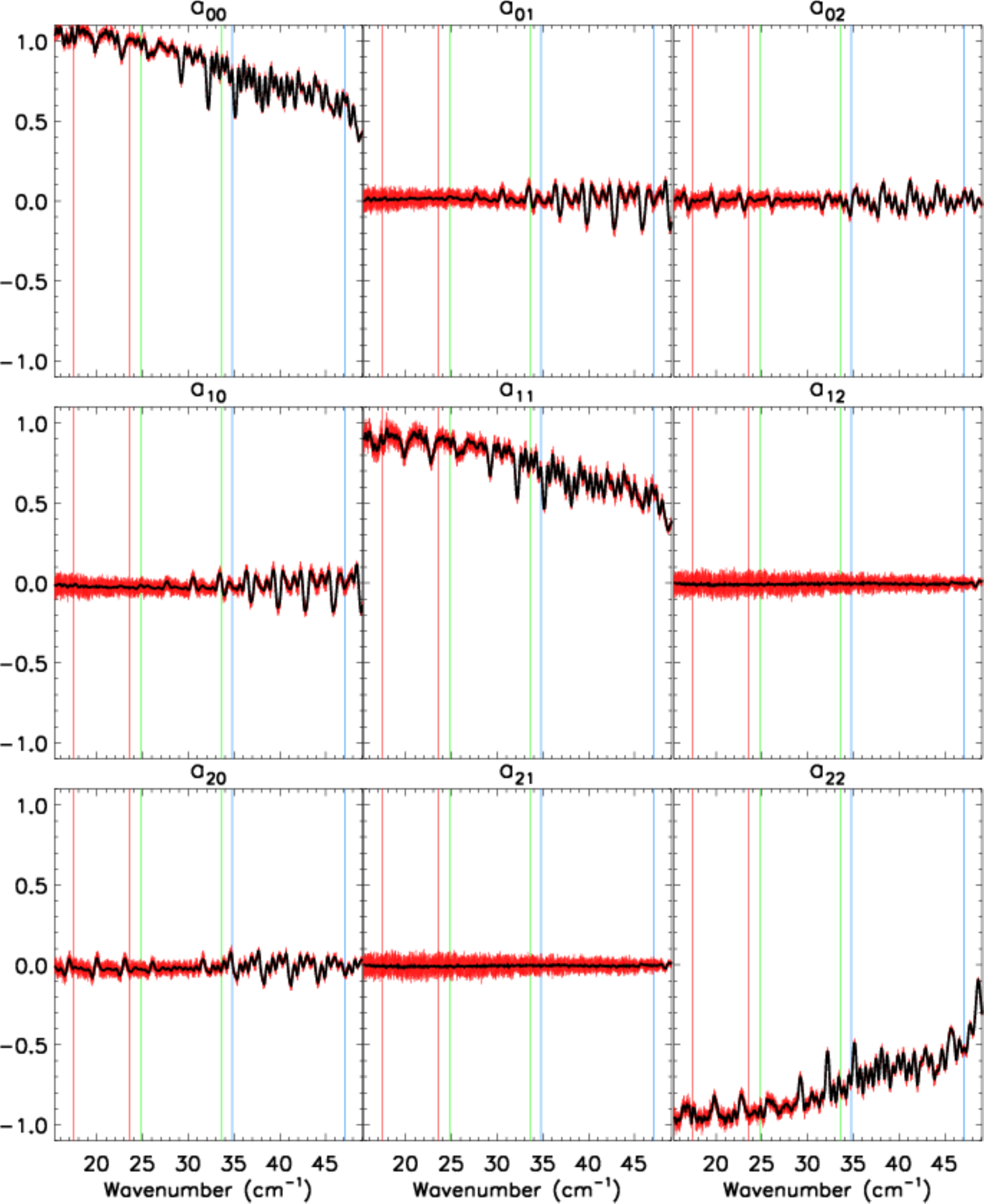}
\caption{Graphical representation of the Mueller matrix of the
BLASTPol HWP, equivalent to that shown in
Fig.~\ref{fig:spare_HWP_MM_coeff_no_offset}, but including in the
fit the frequency-dependent position of the HWP equivalent axes
(see Fig.~\ref{fig:HWP_offset_vs_freq}).}
\label{fig:spare_HWP_MM_coeff}
\end{figure}

Therefore, we include $\beta_{\rm ea}\left(\nu\right)$ in our
Monte Carlo as a frequency-dependent offset in the array of
rotation angles (so that $\theta \rightarrow \theta - \beta_{\rm
ea}$), and repeat our simulations. The results, presented in
Fig.~\ref{fig:spare_HWP_MM_coeff}, can now be qualitatively
compared to the Mueller matrix of an ideal HWP
(Equation~\ref{eq:ideal_HWP}). The improvement is noticeable,
especially in the off-diagonal elements, and the resemblance to an
ideal HWP is remarkable across the entire spectral range of
interest; this procedure effectively acts to diagonalise the HWP
Mueller matrix. However, the transmission losses due to absorption
from the sapphire at $\sim$120\,K still affect the diagonal
elements of the matrix, as expected.

\begin{figure}
\centering
\includegraphics[width=1.\linewidth]{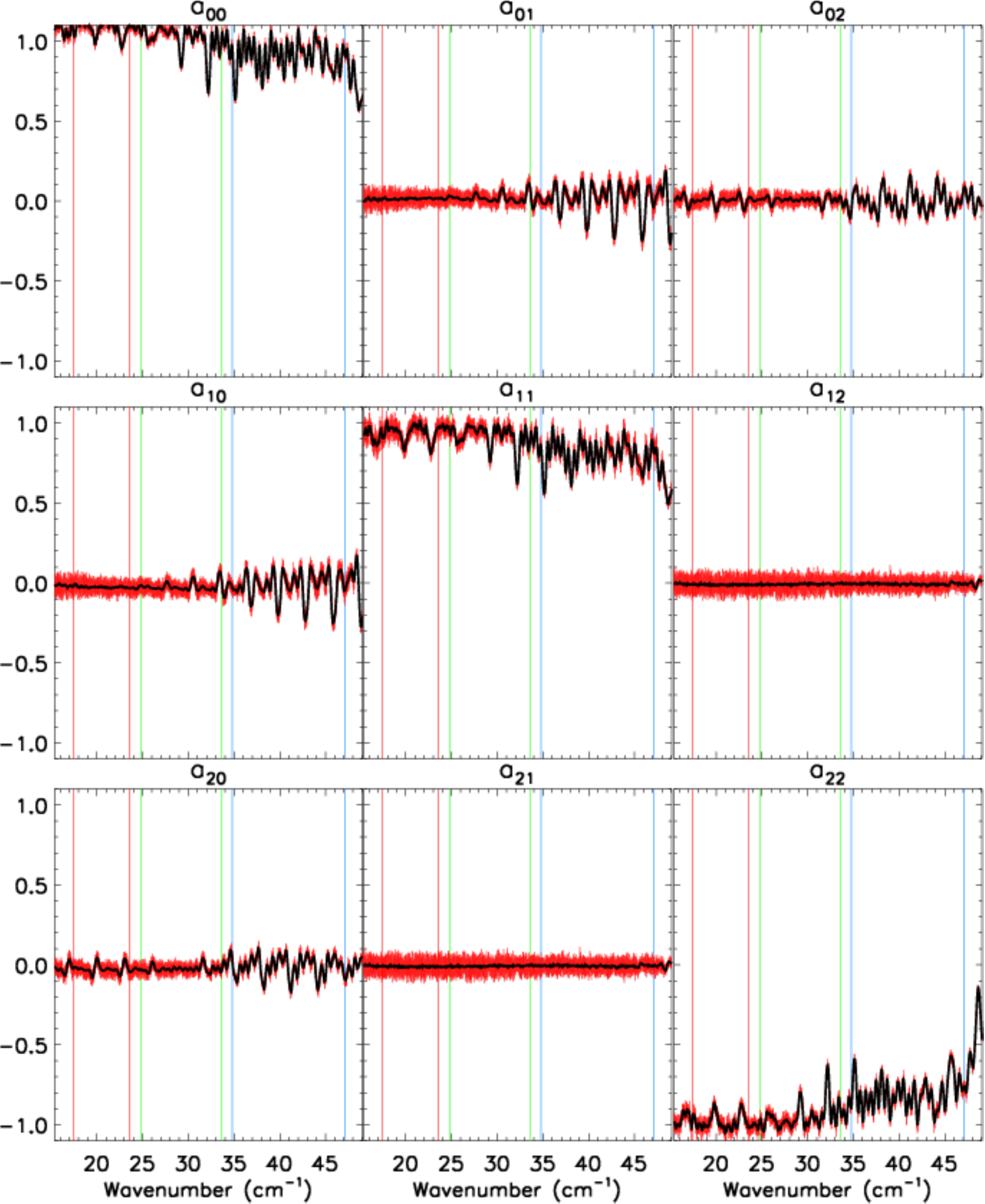}
\caption{Graphical representation of the Mueller matrix of the
BLASTPol HWP, equivalent to that shown in
Fig.~\ref{fig:spare_HWP_MM_coeff}, but including a correction for
the temperature dependence of the sapphire absorption coefficient
(see Fig.~\ref{fig:sapphire_absorption_coefficient_cold}).}
\label{fig:spare_HWP_MM_coeff_cold}
\end{figure}

As a final improvement, we extrapolate the $\beta_{\rm
ea}$-corrected HWP Mueller matrix to 4\,K by including in our
Monte Carlo a correction for the residual sapphire absorption
(using the data presented in
Fig.~\ref{fig:sapphire_absorption_coefficient_cold}). The results
are shown in Fig.~\ref{fig:spare_HWP_MM_coeff_cold}. Although
there still seems to be residual transmission losses due to
sapphire absorption at 250 and 350\,$\mu$m, the retrieved HWP
Mueller matrix is nearly that of an ideal HWP. The band-averaged
values of the matrix coefficients for a flat-spectrum input source
are reported in Table~\ref{tab:spare_HWP_MM_coeff_cold}, along
with their propagated uncertainty; the off-diagonal elements are
always consistent with zero within 2\,$\sigma$ and the modulus of
the three diagonal coefficients is always $>$\,0.8. The
combination of these coefficients with the band-averaged values of
$\beta_{\rm ea}$ given in Table~\ref{tab:HWP_offset_vs_freq} gives
a complete account of the HWP non-idealities to the best of our
ability.

We repeat the calculation of the band-averaged coefficients for
the other spectral indices discussed in
Fig.~\ref{fig:HWP_offset_vs_freq}; we find values that are always
within 1--2\% of those reported in
Table~\ref{tab:spare_HWP_MM_coeff_cold}, and thus we do not
explicitly report them here. Because the three diagonal elements
of the HWP Mueller matrix effectively determine the HWP
co-pol/cross-pol transmission and modulation efficiency, this
analysis confirms that these quantities are very weakly dependent
on the spectral index of the input source; these findings are in
very good agreement with those of \citet{Savini2009}. We will see
in Section~\ref{sec:map_maker} how $\overline{a}_{00}$,
$\overline{a}_{11}$, and $\overline{a}_{22}$ can be incorporated
in the map-making algorithm in terms of optical efficiency,
$\eta$, and polarisation efficiency, $\varepsilon$, of each
detector.

\begin{table}
  \centering
  \begin{tabular}{crrr}
   \hline
   Band & 250\,$\mu$m~~~~ & 350\,$\mu$m~~~~ & 500\,$\mu$m~~~~\\
  \hline
$\overline{a}_{00}$ & 0.905 $\pm$ 0.006 & 1.001 $\pm$ 0.006 & 1.008 $\pm$ 0.007\\
$\overline{a}_{01}$ & 0.012 $\pm$ 0.010 & 0.017 $\pm$ 0.010 & 0.014 $\pm$ 0.011\\
$\overline{a}_{02}$ & $-$0.002 $\pm$ 0.008& 0.006 $\pm$ 0.009 & 0.001 $\pm$ 0.009\\
$\overline{a}_{10}$ & $-$0.016 $\pm$ 0.010 & $-$0.021 $\pm$ 0.010 & $-$0.020 $\pm$ 0.011\\
$\overline{a}_{11}$ & 0.806 $\pm$ 0.011 & 0.928 $\pm$ 0.010 & 0.935 $\pm$ 0.012\\
$\overline{a}_{12}$ & $-$0.007 $\pm$  0.011 &  $-$0.009 $\pm$ 0.014 &  $-$0.011 $\pm$ 0.014\\
$\overline{a}_{20}$ & $-$0.008 $\pm$ 0.008 & $-$0.022 $\pm$ 0.010 & $-$0.021 $\pm$ 0.010\\
$\overline{a}_{21}$ & $-$0.007 $\pm$ 0.011 &  $-$0.009 $\pm$ 0.014 & $-$0.011 $\pm$ 0.014\\
$\overline{a}_{22}$ & $-$0.808 $\pm$ 0.008 & $-$0.960 $\pm$ 0.009 &  $-$0.979 $\pm$ 0.010\\
 \hline
 \end{tabular}
 \caption{Band-averaged Mueller matrix
coefficients. These values are relative to
Fig.~\ref{fig:spare_HWP_MM_coeff_cold}. The input source is
assumed to have a flat spectrum in frequency.}
 \label{tab:spare_HWP_MM_coeff_cold}
\end{table}

\subsection{Effective HWP phase shift}\label{sec:phase_shift}

Finally, we discuss a potential limitation to any linear
polarisation modulator, i.e. the leakage between axes. In a HWP,
the phase shift between the two axes should be as close to
180$^{\circ}$ as possible to avoid transforming linear into elliptical polarisation, hence losing
modulation efficiency. The phase can not be directly measured in a pFTS, but
it can be indirectly inferred from the HWP Mueller matrix.

In order to recover the wavelength-dependent phase shift of the
HWP, we recall the Mueller matrix of a non-ideal impedance-matched
single birefringent slab \citep[][at
$\theta=0^{\circ}$]{Savini2009}:
\begin{equation}\label{eq:mueller_HWP_11}
{\mathbfss M_{\rm
slab}}\left(\theta=0^{\circ},\Delta\varphi\right)=\frac{1}{2}\times~~~~~~~~~~~~~~~~~~~~~~~~~~~~~~~~~~~~~~~~~~~~~~~~~~~~~~~~~
\end{equation}
\begin{equation*}
\nonumber\times\begin{pmatrix}
  \tau_{\|}^2+\tau_{\bot}^2 & \tau_{\|}^2-\tau_{\bot}^2 & 0 & 0 \\
  \tau_{\|}^2-\tau_{\bot}^2 & \tau_{\|}^2+\tau_{\bot}^2 & 0 & 0 \\
  0 & 0 & 2\,\tau_{\|}\,\tau_{\bot}\cos\Delta\varphi & 2\,\tau_{\|}\,\tau_{\bot}\sin\Delta\varphi \\
  0 & 0 & -2\,\tau_{\|}\,\tau_{\bot}\sin\Delta\varphi & 2\,\tau_{\|}\,\tau_{\bot}\cos\Delta\varphi \\
\end{pmatrix}~,~~~~~~~~~~~~~~~~~
\end{equation*}
where $\tau_{\|}^2$ and $\tau_{\bot}^2$ are the measured transmissions of
orthogonal polarisations aligned with the birefringent
axes. By comparing the matrix in Equation~\ref{eq:mueller_HWP_11} with
that of a generic HWP, we can solve for the HWP phase shift as
\begin{equation}\label{eq:mueller_HWP_12}
\cos\Delta\varphi =\frac{a_{22}}{2}
\left({\frac{a_{00}+a_{01}}{2}}\right)^{-\frac{1}{2}}
\left({\frac{a_{00}-a_{01}}{2}}\right)^{-\frac{1}{2}}~.
\end{equation}

\begin{figure}
\centering
\includegraphics[width=1.\linewidth]{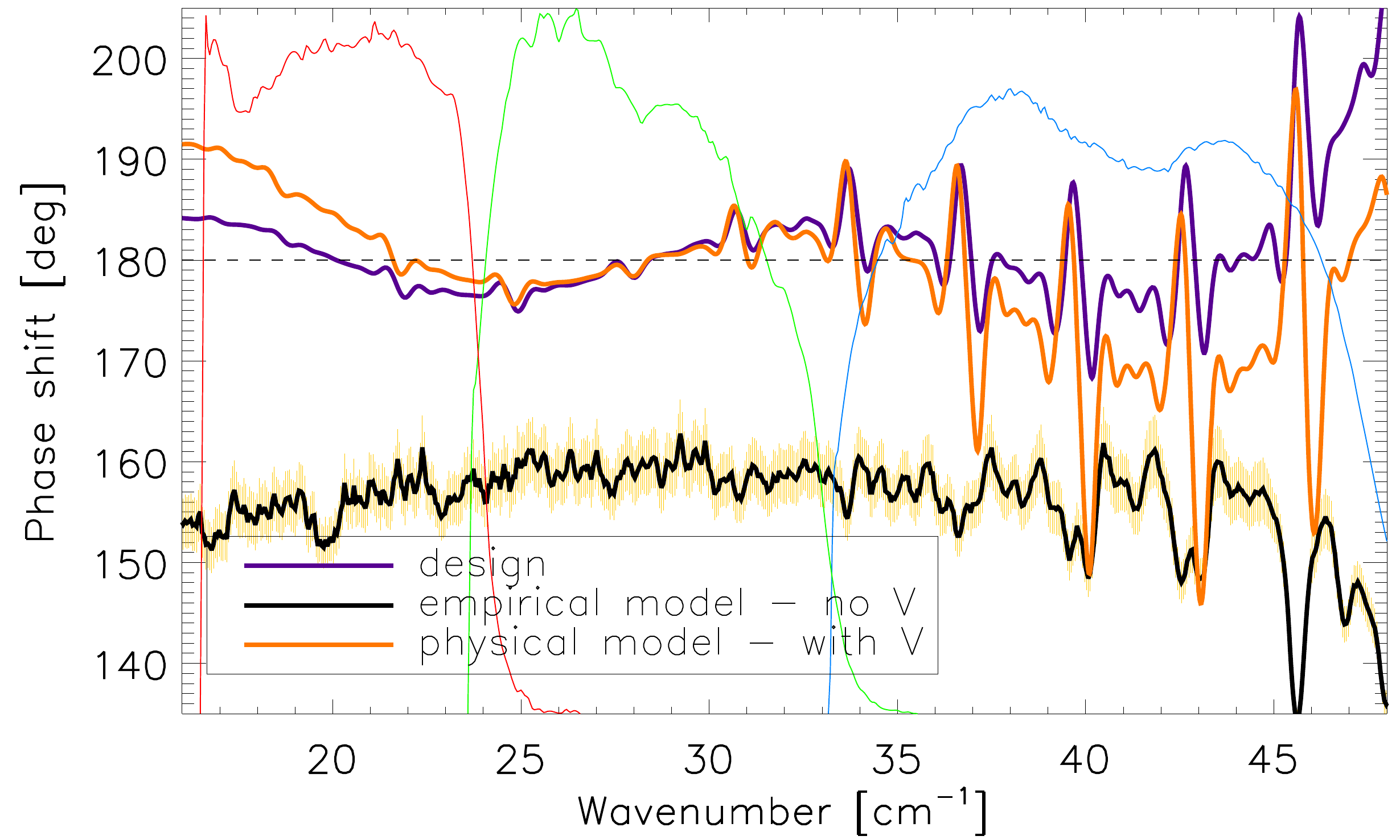}
\caption{Phase shift of the BLASTPol HWP as a function of wavenumber. The purple line shows the design goal and the orange line shows the {\it as-built} phase shift, estimated using the physical model of \citet[][see Section~\ref{sec:physical_model}]{Savini2006}, which accounts for the depolarisation effects due to standing waves between sapphire substrates by estimating the circular polarisation ($V$) portion of the HWP Mueller matrix. On the other hand, our empirical model (in black, with 3\,$\sigma$ error bars in yellow) underestimates the phase shift because (in the absence of the $V$ terms) we have assumed a pure cosine modulation, which is not adequate in a multi-slab Pancharatnam stack. Also shown for reference are the ideal phase shift for a HWP (180$^{\circ}$) and the relative spectral response of the three BLASTPol channels, in arbitrary units.} \label{fig:phase_spareHWP}
\end{figure}

Equation~\ref{eq:mueller_HWP_12} allows us to recover the phase
shift from our knowledge of $a_{00}$, $a_{01}$ and $a_{22}$.
Fig.~\ref{fig:phase_spareHWP} shows the estimated phase shift of
the BLASTPol HWP as a function of wavenumber, after the
introduction in our Monte Carlo routine of the
wavelength-dependent position of the HWP equivalent axes depicted
in Fig.~\ref{fig:HWP_offset_vs_freq}.


Clearly, the $\beta_{\rm ea}$-corrected phase shift (in black) appreciably departs from 180$^{\circ}$. However, recall that Equation~\ref{eq:mueller_HWP_12}
strictly applies only to a single birefringent plate. In a
multi-slab Pancharatnam stack, $\Delta\varphi$ becomes an
``effective'' phase shift as we are no longer in the presence of a
pure cosine modulation, thus slightly skewing the HWP modulation
function and resulting in an artificially higher leakage between
axes when the cosine function is inverted.

Our empirical method, which does not estimate the circular polarisation ($V$) portion of the HWP Mueller matrix, is prone to underestimating the phase shift by mistaking the depolarisation effects due to standing waves between sapphire substrates (enhanced by the presence of the interspersed polyethylene layers) for a phase deficit. In fact, were the phase shift really $\lesssim$\,160$^{\circ}$, we would get a substantially higher cross-pol on the FTS measurements than the $\la$\,0.5\% we measure in Figs.~\ref{fig:BLAST_pol_spareHWP_ARC_xpol_spectra} and \ref{fig:modeff}a.

While none of the other quantities estimated in the previous sections are affected by this deficiency in our model, we redress for the inadequacy of our empirical model in retrieving the phase shift by resorting to the physical model of \citet[][discussed in Section~\ref{sec:physical_model}]{Savini2006}. The purple line in Fig.~\ref{fig:phase_spareHWP} shows the design goal for the BLASTPol HWP, which is computed by assuming the nominal values for the HWP build parameters (see Table~\ref{tab:physical_parameters_HWP}), while the orange line shows the as-built phase shift, which is obtained by fitting the spectral data and allowing the build parameters to vary in a physical way around the nominal values.

The physical model is able to reconstruct the inevitable substrate alignment errors that occur during the HWP assembly and provides a much improved estimate of the actual phase shift, which is within $\sim$\,5$^{\circ}$ of the ideal value for the central BLASTPol band (350\,$\mu$m, where the HWP is optimized). Although the side bands show worse performance (within $\sim$\,15$^{\circ}$ of 180$^{\circ}$), we have indications that the modulation efficiency of the HWP at 4\,K is only mildly affected by this departure from
ideality. From Fig.~\ref{fig:modeff}b we see that the extrapolated
HWP modulation efficiency is always above 95\% across the whole
spectral range of interest, with band-integrated values exceeding
98\%. Moreover, phase shift deviations of similar amplitude are
measured in most (sub)mm-wave achromatic half-wave plates
manufactured to date \citep[e.g.,][]{Savini2009,Zhang2011}.

In addition to the non-ideal substrate alignment, the physical model yields a best-fit thickness for the individual sapphire substrates that is 9.6\,$\mu$m smaller than the nominal value of 500\,$\mu$m available on the market\footnote{We verify that the sum of the thicknesses from Table~\ref{tab:physical_parameters_HWP} is compatible with the overall measured thickness of the HWP stack.}. While it is beyond the scope of this paper to quantify exactly the relative contributions of plate misalignment and reduced thickness to the overall non-ideal behaviour of the phase shift, we can not rule out a contribution from both effects. Furthermore, and more importantly, neither of these two effects has nearly the same impact on the other performance parameters presented in the previous sections (the 9 Mueller matrix coefficients associated with linear polarisation) because of their weak dependence on the phase shift.

Incidentally, we verify that our methodology does not violate
conservation of energy by ensuring that the output Stokes vector
resulting from a generic polarised input travelling through the
recovered HWP Mueller matrix satisfies $I^2 \geq Q^2 + U^2$ in
every instance described above.

\section{Map-making algorithm}\label{sec:map_maker}

Map-making is the operation that generates an astronomical map,
which contains in every pixel an estimate of the sky emission, and
is obtained by combining data from all detectors available at a
given wavelength channel, their noise properties and the pointing
information. The raw BLAST(Pol) data consist of bolometer time-ordered
streams, which are cleaned and pre-processed before
being fed into the map-maker; the details are extensively described elsewhere
\citep{Rex2007,Truch2007,Wiebe2008,Pascale2008}, and we refer to
these works for a complete account of the low-level data
reduction.

In the following, we focus on the mathematical formalism of the
map-making technique, and its algorithmic implementation in the
specific case of BLASTPol.


For a non-ideal polarisation experiment, by adopting the Stokes
formalism and assuming that no circular ($V$) polarisation is
present, we can model the data as
\begin{equation}\label{eq:timeline}
    d^i_t = \frac{\eta^i}{2}A^i_{tp}\left[ I_p + \varepsilon^i \left(Q_p\,\cos2\gamma^i_t + U_p\,\sin 2\gamma^i_t \right)\right] + n^i_t~.
\end{equation}
Here: $i$, $t$ and $p$ label detector index, time, and map pixel
respectively; $d^i_t$ are the time-ordered data for a given
channel, related to the sky maps $\left[I_p,Q_p,U_p\right]$ by the
pointing operator $A^i_{tp}$; $\eta^i$ is the optical efficiency
of each detector; $\varepsilon^i$ is the polarisation efficiency
of each detector with its polarising grid (analyser); and $n^i_t$
represents a generic time-dependent noise term. Throughout this
discussion it is assumed that the term within square brackets is
the convolution of the sky emission with the telescope
point-spread function (PSF). $\gamma^i_t$ is the time-ordered
vector of the observed polarisation angle, defined as the angle
between the polarisation reference vector at the sky pixel $p$ (in
the chosen celestial frame) and the polarimeter transmission axis.
$\gamma^i_t$ is given by
\begin{equation}\label{eq:polangle}
    \gamma^i_t = \alpha_t^i + 2\left[\beta_t-\beta_0-\overline{\beta}_{\rm ea}\right] +\delta^i_{\rm grid}~,
\end{equation}
where $\alpha_t^i$ is the angle between the reference vector at
pixel $p$ and a vector pointing from $p$ to the zenith along a
great circle, $\beta_t$ is the HWP orientation angle in the
instrument frame, $\beta_0$ is the HWP zero angle in the
instrument frame, $\overline{\beta}_{\rm ea}$ is the band-averaged
position of the equivalent axes of the HWP (dependent on the known
or assumed spectral signature of the input source; see
Section~\ref{sec:beta_ea}), and $\delta^i_{\rm grid}=[0,\pi/2]$
accounts for the transmission axis of the polarising grids being
parallel/perpendicular to the zenith angle.

The notation outlined above can be connected to the Mueller
formalism developed in Section~\ref{sec:mueller_HWP} to determine
under which circumstances Equation~\ref{eq:timeline} is valid in
the presence of a real (i.e., non-ideal) HWP. Because we have
included in Equation~\ref{eq:polangle} the band-averaged position
of the equivalent axes of the HWP, $\overline{\beta}_{\rm ea}$,
the Mueller matrix of the BLASTPol HWP can be considered almost
that of an ideal HWP, as discussed in Section~\ref{sec:beta_ea}.
Nonetheless, we have shown that the band-averaged values of the
three diagonal matrix coefficients are not identically unity (but
always $>$\,0.8 in modulus), probably as a result of residual
absorption from sapphire, especially in the 250 and 350\,$\mu$m
bands (although we have corrected for it to the best of our
knowledge).

In the light of these considerations, we now want to compare
Equation~\ref{eq:timeline} to Equation~\ref{eq:mueller_HWP_4b},
which both represent the signal measured by a polarisation
insensitive intensity detector when illuminated by a polarised
input that propagates through a rotating HWP and an analyser. A
term-by-term comparison shows that these two expressions are
equivalent when the coefficients $B$ and $C$ (defined in
Equation~\ref{eq:mueller_HWP_5}) are zero, i.e. when the HWP
modulates the polarisation purely at four times the rotation
angle, with no leakage in the second harmonic (twice the rotation
angle) and thus no leakage of $I$ into $Q$ and $U$. These two
coefficients are linear combinations of the HWP Mueller matrix
elements $\overline{a}_{01}, \overline{a}_{10}, \overline{a}_{02},
\overline{a}_{20}$, which we have shown in
Table~\ref{tab:spare_HWP_MM_coeff_cold} to be all compatible with
zero within 2\,$\sigma$. In addition, their amplitude is at most
$\sim$2\% of that of the diagonal matrix elements, 
so to first order the coefficients $B$ and $C$ can be
neglected, and the two expressions can be
considered equivalent. Nonetheless, these generally moderate
levels of $I\rightarrow Q,U$ leakage can be readily accounted for
by incorporating in the map-making algorithm a correction for the
``instrumental polarisation'' \citep[IP; see ][and Angil\`{e} et al. in preparation]{Matthews2013}.

In addition, after some simple algebra, it can be shown that
$\eta={a}_{00}+\frac{{a}_{11}}{2}+\frac{{a}_{22}}{2}$ and
$\eta\,\varepsilon=\frac{{a}_{11}}{2}-\frac{{a}_{22}}{2}$. As
anticipated in Section~\ref{sec:beta_ea}, the knowledge of the
band-averaged values of the three diagonal matrix elements,
$\overline{a}_{00}, \overline{a}_{11}, \overline{a}_{22}$ (which
we have shown to depend weakly on the spectral index of the input
source), can be readily incorporated in the map-making algorithm
in terms of optical efficiency, $\eta$, and polarisation
efficiency, $\varepsilon$, of the HWP; these can be factored into
the overall optical and polarisation efficiency of each
detector, which are the product of several factors (e.g., bolometer absorption efficiency, HWP efficiency, absorption in the optical chain, etc.). From the values listed in Table~\ref{tab:spare_HWP_MM_coeff_cold}, in our case we find
$[\eta_{\rm hwp},\varepsilon_{\rm hwp}]=[0.904,0.893],
[0.985,0.958]$, and $[0.986,0.971]$ at 250, 350, and 500\,$\mu$m,
respectively.

Finally, the comparison of Equations~\ref{eq:timeline} and
Equation~\ref{eq:mueller_HWP_4b} also yields the relation $\eta\,\varepsilon\,\chi=-{a}_{12}=-{a}_{21}$, where we have
introduced a new parameter, $\chi$, which quantifies the amplitude
of the mixing of $Q$ and $U$. From
Table~\ref{tab:spare_HWP_MM_coeff_cold}, we see that
$\overline{a}_{12}=\overline{a}_{21}$ are always compatible with
zero within 1\,$\sigma$, and their amplitude is at most $\sim$1\%
of that of the diagonal matrix elements. Nonetheless we quantify
the amplitude of the $Q \leftrightarrow U$ mixing to be $\chi_{\rm
hwp}=0.009,0.010$, and 0.011 at 250, 350, and 500\,$\mu$m,
respectively. While this correction is not currently included in
our algorithm, we indicate that it can be implemented in a
relatively straightforward way by modifying
Equation~\ref{eq:timeline} with a double change of variable, i.e.
$Q\rightarrow Q+\chi U$ and $U\rightarrow U+\chi Q$. If $\chi$ is
estimated to the required accuracy, the unmixed $Q$ and $U$ can be
retrieved unbiasedly. This correction may be very relevant to CMB
polarisation experiments, where any $Q \leftrightarrow U$ leakage
leads to a spurious mixing of the $EE$ and $BB$ modes.

We remind the reader that the above factors have been computed
directly from the band-averaged coefficients of the inferred HWP
Mueller matrix extrapolated to 4\,K, and offer a direct way to
include the modelled HWP non-idealities in a map-making algorithm.
On the other hand, the band-averaged HWP maximum transmission,
polarisation efficiency and cross-pol quoted at the end of
Section~\ref{sec:HWP_spectra} are estimated directly from the
spectra extrapolated to 4\,K, and are mostly informative from an
experimental point of view rather than for data analysis purposes.

Consider now one map pixel $p$ that is observed in one band by $k$
detectors ($i=1,...,k$); let us define the generalised pointing
matrix {\mathbfss A}$_{tp}$, which includes the trigonometric
functions along with the efficiencies,
\begin{equation}\label{eq:generalized_pointing_matrix}
{\mathbfss A}_{tp} \equiv \frac{1}{2}
\begin{pmatrix}
   \eta^1\,A^1_{tp} & \eta^1\,\varepsilon^1\,A^1_{tp}\,\cos2\gamma^1_t & \eta^1\,\varepsilon^1\,A^1_{tp}\,\sin2\gamma^1_t \\
  \vdots & \vdots & \vdots \\
  \eta^i\,A^i_{tp} & \eta^i\,\varepsilon^i\,A^i_{tp}\,\cos2\gamma^i_t & \eta^i\,\varepsilon^i\,A^i_{tp}\,\sin2\gamma^i_t \\
  \vdots & \vdots & \vdots \\
  \eta^k\,A^k_{tp} & \eta^k\,\varepsilon^k\,A^k_{tp}\,\cos2\gamma^k_t & \eta^k\,\varepsilon^k\,A^k_{tp}\,\sin2\gamma^k_t \\
\end{pmatrix}~,
\end{equation}
and the map triplet ${\mathbfss S}_{p}$, along with the combined
detector (${\mathbfit D}_t$) and noise (${\mathbfit n}_t$)
timelines,
\begin{equation}\label{eq:map_triplet}
{\mathbfss S}_{p} \equiv
\begin{pmatrix}
   I_p\\
   Q_p\\
   U_p\\
\end{pmatrix}~,~~~~
{\mathbfit D}_t \equiv
\begin{pmatrix}
   d^1_t\\
  \vdots\\
  d^i_t\\
  \vdots\\
  d^k_t\\
\end{pmatrix}~,~~~~
{\mathbfit n}_t \equiv
\begin{pmatrix}
   n^1_t\\
  \vdots\\
  n^i_t\\
  \vdots\\
  n^k_t\\
\end{pmatrix}~.
\end{equation}
Equation~\ref{eq:timeline} can then be rewritten in compact form
as
\begin{equation}\label{eq:timeline_matrix}
    {\mathbfit D}_t = {\mathbfss A}_{tp}\,{\mathbfss S}_{p} + {\mathbfit n}_t~.
\end{equation}

Under the assumption that the noise is Gaussian and stationary,
the likelihood of {\mathbfss S}$_{p}$ given the data can be
maximised, thus yielding the well-known generalised least squares
(GLS) estimator for {\mathbfss S}$_{p}$:
\begin{equation}\label{eq:GLS}
    {\tilde{\mathbfss S}}_{p} = \left({\mathbfss A}_{tp}^{\rm T}\,{\mathbfss N}^{-1}\,{\mathbfss A}_{tp}\right)^{-1}\,{\mathbfss A}_{tp}^{\rm T}\,{\mathbfss
    N}^{-1}\,{\mathbfit D}_t~,
\end{equation}
where ${\mathbfss N} \equiv \langle{\mathbfit n}_t\,{\mathbfit
n}_{t'}\rangle$ is the noise covariance matrix of the data in
the
time domain, with $t,t'$ running over the detector time samples (typically
$N_{\rm s}\sim10^6$--10$^7$).


Computation of the solution to Equation~\ref{eq:GLS} is far from
trivial in most astronomical applications, due to {\mathbfss N}
being a very large matrix, of size $k N_{\rm s} \times k N_{\rm
s}$. Understandably, it is computationally challenging to invert
this matrix, especially when there are correlations among
detectors, and a number of ``optimal'' map-making techniques have
been developed in the literature to tackle this problem
\citep[e.g.,][]{Natoli2001,Natoli2009,Masi2006,Johnson2007,Wu2007,Patanchon2008,Cantalupo2010}.

\subsection{Naive binning}

In the simple case that the noise is uncorrelated between
different detectors, the matrix {\mathbfss N} reduces to block diagonal:
\begin{equation}\label{eq:noise_covariance_diag}
    \langle n^i_t\,n^j_{t'}\rangle = \langle n^j_t\,n^i_{t'}\rangle = 0~~~~~(i\neq j)~.
\end{equation}
In addition, let us assume that there is no correlation between
noise of different samples acquired by the same detector, or, in
other words, that the noise in each detector is white. In this case, each ``block''
of the noise covariance matrix collapses into one value, which is
the timeline variance for each detector. Hence, {\mathbfss N}
becomes a $k \times k$ diagonal matrix where the diagonal elements
are the sample variances of the detectors, $\sigma^2_i$, and
weights can thus be defined as the inverse of those variances,
$w^i \equiv 1/\sigma^2_i$.

Therefore, under the assumption that the noise is white and
uncorrelated among detectors, Equation~\eqref{eq:GLS} reduces to a
simple, weighted binning \citep[``naive'' binning; see
also][]{Benoit2004,Pascale2011} of the map:
\begin{equation}\label{eq:naive_binning}
{\mathbfss S}_{p} =
\begin{pmatrix}
 I_p\\
 Q_p\\
 U_p\\
\end{pmatrix}
= \frac{\sum\limits_{i=1}^{k} \sum\limits_{t=1}^{N_{\rm s}} w^i
\frac{({\mathbfss A}^i_{tp})^{\rm T}\,d^i_t}{({\mathbfss
A}^i_{tp})^{\rm T}\,{\mathbfss A}^i_{tp}}}{\sum\limits_{i=1}^{k}
w^i}~.
\end{equation}

In the light of these considerations, let us go back to
Equation~\eqref{eq:timeline} and model the generic time-dependent
noise term as $n^i_t = u_t + \xi^i \rho_t$~, where $u_t$ represents a time-dependent noise term, completely
uncorrelated among different detectors, while $\rho_t$ describes
the correlated noise (varying over timescales larger than the
ratio of the size of the detector array to the scan speed),
coupled to each detector via the $\xi^i$ parameter, peculiar to
each bolometer.

Let us define the following quantity for every pixel $p$ in the
map:
\begin{equation}\label{eq:linear_system1}
{\mathbfss S}_{p}^{\rm e} =
\begin{pmatrix}
 I_p^{\rm e}\\
 Q_p^{\rm e}\\
 U_p^{\rm e}\\
\end{pmatrix}
\equiv
\begin{pmatrix}
  \sum\limits_{i=1}^{k} \sum\limits_{t=1}^{N_{\rm s}}d^i_t \\
  \sum\limits_{i=1}^{k} \sum\limits_{t=1}^{N_{\rm s}}d^i_t\,\cos2\gamma^i_t\\
  \sum\limits_{i=1}^{k} \sum\limits_{t=1}^{N_{\rm s}}d^i_t\,\sin2\gamma^i_t \\
\end{pmatrix}~,
\end{equation}
where $N_{\rm s}$ is now the number of samples in each detector
timeline that fall within pixel $p$, and the superscript ``e''
stands for ``estimated''. The above quantities can be computed
directly from the detector timelines. Recalling
Equation~\eqref{eq:timeline}, we
can outline the following linear system of $3$ equations with $3$
unknowns:
\begin{align}\label{eq:linear_system2}
\nonumber\begin{pmatrix}
 I_p^{\rm e}\\
 Q_p^{\rm e}\\
 U_p^{\rm e}\\
\end{pmatrix}
=
\begin{pmatrix}
  \sum\limits_{i,t} 1 & \sum\limits_{i,t} \cos2\gamma^i_t & \sum\limits_{i,t} \sin2\gamma^i_t \\
  \sum\limits_{i,t} \cos2\gamma^i_t & \sum\limits_{i,t} \cos^2 2\gamma^i_t & \sum\limits_{i,t} \cos2\gamma^i_t\,\sin2\gamma^i_t \\
  \sum\limits_{i,t} \sin2\gamma^i_t & \sum\limits_{i,t} \cos2\gamma^i_t\,\sin2\gamma^i_t & \sum\limits_{i,t} \sin^2 2\gamma^i_t \\
\end{pmatrix}\\
\cdot  \frac{1}{2}
\begin{pmatrix}
 I_p\\
 Q_p\\
 U_p\\
\end{pmatrix}
+
\begin{pmatrix}
  \sum\limits_{i,t} (u_t + \xi^i \rho_t)\\
  \sum\limits_{i,t} (u_t + \xi^i \rho_t)\,\cos2\gamma^i_t\\
  \sum\limits_{i,t} (u_t + \xi^i \rho_t)\,\sin2\gamma^i_t\\
\end{pmatrix}~,~~~~~~~~~~~~~~~~~~~
\end{align}
where we have temporarily assumed $\eta^i=\varepsilon^i=w^i=1$ and
combined the two sums in one, with the indices $i$ and $t$
running, respectively, over the bolometers and the samples in each
detector timeline.

If we now define the quantities,
\begin{eqnarray}\label{eq:linear_system3}
\nonumber N_{\rm hit}\equiv \sum_{i,t} \frac{1}{2},~~~~~~c
\equiv \sum_{i,t} \frac{1}{2} \cos2\gamma^i_t,\\
\nonumber c_2 \equiv\sum_{i,t} \frac{1}{2} \cos^2
2\gamma^i_t,~~~~~~
s \equiv \sum_{i,t} \frac{1}{2} \sin2\gamma^i_t,\\
\nonumber s_2\equiv\sum_{i,t} \frac{1}{2} \sin^2 2\gamma^i_t =
N_{\rm hit}-c_2,~~~~~~U\equiv\sum_{i,t} u_t,\\
\nonumber m\equiv\sum_{i,t} \frac{1}{2}
\cos2\gamma^i_t\,\sin2\gamma^i_t
\nonumber,~~~~C^u_2\equiv\sum_{i,t} u_t\,\cos2\gamma^i_t\\
\nonumber S^u_2\equiv\sum_{i,t} u_t\,\sin2\gamma^i_t,~~~~~~
P\equiv\sum_{i,t} \xi^i \rho_t,\\
C^{\rho}_2\equiv\sum_{i,t} \xi^i
\rho_t\,\cos2\gamma^i_t,~~~~~~S^{\rho}_2\equiv\sum_{i,t} \xi^i
\rho_t\,\sin2\gamma^i_t,
\end{eqnarray}
then the system in Equation~\eqref{eq:linear_system2} can be
rewritten in compact form as
\begin{equation}\label{eq:linear_system4}
\begin{pmatrix}
 I_p^{\rm e}\\
 Q_p^{\rm e}\\
 U_p^{\rm e}\\
\end{pmatrix}
=
\begin{pmatrix}
  N_{\rm hit} & c & s \\
  c & c_2 & m \\
  s & m & N_{\rm hit}-c_2 \\
\end{pmatrix}
\cdot
\begin{pmatrix}
 I_p\\
 Q_p\\
 U_p\\
\end{pmatrix}
+
\begin{pmatrix}
  U+P\\
  C^u_2 + C^{\rho}_2\\
  S^u_2 + S^{\rho}_2\\
\end{pmatrix}~.
\end{equation}

In order to retrieve an estimate of ${\mathbfss S}_{p}$ from the
quantities computed in Equation~\eqref{eq:linear_system1}, the
above system has to be solved for every pixel $p$ in the map. One
can already see the computational advantage of inverting a
$3\times3$ matrix $N_{\rm pix}\times N_{\rm pix}$ times, with
respect the inversion of a generic $k N_{\rm s} \times k N_{\rm
s}$ matrix \citep[for detectors having uncorrelated $1/f$ noise as
well as a common-mode $1/f$ noise;][]{Patanchon2008}, or $k$
matrices of size $N_{\rm s} \times N_{\rm s}$ \citep[for detectors
having only uncorrelated $1/f$ noise;][]{Cantalupo2010}. The main
difficulties are, of course, in estimating the noise terms
$U,P,C^u_2,C^{\rho}_2,S^u_2,S^{\rho}_2$. However, recalling
Equation~\eqref{eq:polangle} and the fact that adjacent detectors
have orthogonal polarising grids ($\delta^i_{\rm
grid}=[0,\pi/2]$), we note that, in the sum over $i$, adjacent
detectors have equal and opposite contributions to $C^{\rho}_2$
and $S^{\rho}_2$, under the following assumptions: (i) the
timescale over which the correlated noise is approximately
constant is larger than the time elapsed while scanning the same
patch of sky with two adjacent detectors, and (ii) $\xi^i$ is not
too dissimilar between adjacent bolometers.

This means that the terms $C^{\rho}_2$ and $S^{\rho}_2$ can be
neglected, under the above assumptions, while estimating the
$[Q,U]$ maps. In particular, as a first step, we can solve for $I$
only by high-pass filtering the timelines, in order to suppress
the correlated noise term in $I$, $P$. Subsequently, $I$ can be
assumed known, and the $[Q,U]$ maps can be computed without
filtering the timelines, so that polarised signal at large angular
scales is not suppressed.


The other assumption required for the naive binning is that the
noise is white, at least on the timescales relevant to BLASTPol's
scan strategy. An analysis of the bolometer timelines from the
2010 campaign shows that the knee of the $1/f$ noise in the
difference between two adjacent detectors is typically located at
frequencies $\lesssim$\,0.1\,Hz; assuming a typical scan speed of
0.1$^{\circ}$\,s$^{-1}$, this corresponds to angular scales of
$\gtrsim$\,1$^{\circ}$ on the sky. The regions mapped by BLASTPol
hardly exceed 1$^{\circ}$ in size \citep[see Table 1.1
in][and Angil\`{e} et al. in preparation]{Moncelsi2011b}, hence here we stipulate that the noise in
the difference between pairs of adjacent detectors is white.

Therefore, under the assumptions above, we can solve the linear
system outlined in Equation~\eqref{eq:linear_system4}; by defining
the following quantities,
\begin{eqnarray}\label{eq:linear_system5}
\nonumber \Delta & \equiv & c^2\left(c_2-N_{\rm h}\right)-N_{\rm h}\left(c_2^2+m^2-c_2\,N_{\rm h}\right)+2\,c\,s\,m-c_2\,s^2~, \\
\nonumber A &\equiv&-\left(c_2^2+m^2-c_2\,N_{\rm h}\right),~~~~B\equiv c\left(c_2-N_{\rm h}\right)+s\,m~,\\
 C &\equiv& c\,m-s\,c_2 ,~~~~D\equiv -\left[\left(c_2-N_{\rm h}\right)N_{\rm h}+s^2\right]~,\\
\nonumber E &\equiv& c\,s-m\,N_{\rm h},~~~~F\equiv c_2\,N_{\rm
h}-c^2~,
\end{eqnarray}
the solution to the system can be written in compact form:
\begin{equation}\label{eq:linear_system6}
{\mathbfss S}_{p} =
\begin{pmatrix}
 I_p\\
 Q_p\\
 U_p\\
\end{pmatrix}
=
\begin{pmatrix}
\frac{A\,I^{\rm e}_p+B\,Q^{\rm e}_p+C\,U^{\rm e}_p}{\Delta}\\
\frac{B\,I^{\rm e}_p+D\,Q^{\rm e}_p+E\,U^{\rm e}_p}{\Delta}\\
\frac{C\,I^{\rm e}_p+E\,Q^{\rm e}_p+F\,U^{\rm e}_p}{\Delta}\\
\end{pmatrix}~,
\end{equation}
where we have renamed $N_{\rm hit} \rightarrow N_{\rm h}$ for
brevity.

\subsection{Weights and uncertainties}

The solution for ${\mathbfss S}_{p}$ given in
Equation~\eqref{eq:linear_system6} is a simple, unweighted binning
of the data into the map pixels. In reality, as anticipated in
Equation~\eqref{eq:naive_binning}, we want to perform a weighted
binning, where the weight of each detector is given by the inverse
of its timeline variance, which can be easily measured as the
bolometer's white noise floor level. In our formalism, the
weighted binning is simply achieved by defining $[I^{\rm
e}_p,Q^{\rm e}_p,U^{\rm e}_p]$ in
Equation~\eqref{eq:linear_system1}, as well as each of the
quantities $N_{\rm h}$, $c$, $s$, $c_2$, $s_2$, and $m$ introduced
in Equation~\eqref{eq:linear_system3}, to include $w^i$ in the
sums. Similarly, the measured values of the optical efficiencies
$\eta^i$ and polarisation efficiencies $\varepsilon^i$ can readily
be inserted in Equation~\eqref{eq:linear_system2} to account for
the non-idealities of the optical system.

The introduction of the weights allows us to derive the expression
for the statistical error on ${\mathbfss S}_{p}$, in the continued
assumption of uncorrelated noise, following the usual error
propagation formula \citep[e.g.,][here we omit the sum over $t$
for simplicity]{NumericalRecipes}:
\begin{equation}\label{eq:map_maker_uncert1}
\sigma^2_p = \sum_{i} \frac{1}{w^i}\left(\frac{\partial {\mathbfss
S}_{p}}{\partial d^i}\right)^2~.
\end{equation}
After some tedious algebra, the expression for the statistical
error is
\begin{eqnarray}\label{eq:map_maker_uncert2}
\nonumber\sigma^2_{p} =
\begin{pmatrix}
 {\rm Var}^I_p\\
 {\rm Var}^Q_p\\
 {\rm Var}^U_p\\
\end{pmatrix}
=~~~~~~~~~~~~~~~~~~~~~~~~~~~~~~~~~~~~~~~~~~~~~~~~~~~~~~~~~~~~\\
\nonumber
\begin{pmatrix}
\frac{2}{\Delta^2}\left(A^2\,N_{\rm h}+B^2\,c_2+C^2\,s_2+2\,A\,B\,c+2\,A\,C\,s+2\,B\,C\,m\right)\\
\frac{2}{\Delta^2}\left(B^2\,N_{\rm h}+D^2\,c_2+E^2\,s_2+2\,B\,D\,c+2\,B\,E\,s+2\,D\,E\,m\right)\\
\frac{2}{\Delta^2}\left(C^2\,N_{\rm h}+E^2\,c_2+F^2\,s_2+2\,C\,E\,c+2\,C\,F\,s+2\,E\,F\,m\right)\\
\end{pmatrix},
\end{eqnarray}
where $s_2\equiv N_{\rm h}-c_2$, as noted in
Equation~\eqref{eq:linear_system3}. To first order, these
expressions can be used to quantify the uncertainty of $[I,Q,U]$ in
each map pixel $p$. A more comprehensive account of the
correlations in the noise, as well as a thorough validation of the
assumptions made here, is beyond the scope of this paper and will
be treated in a future work.

\subsection{Preliminary map}\label{sec:map}

As a proof of concept of the naive binning technique for the
BLASTPol polarised map-maker, which includes all the corrections
due to HWP non-idealities discussed in this work, we present a
\emph{preliminary} map at 500\,$\mu$m of one of the giant
molecular clouds observed by BLASTPol, the Carina Nebula. In this
specific case, we calculate $\overline{\beta}^{\rm 500\,\mu
m}_{\rm ea} = 1.7$ based on a dust emissivity spectral index of
1.37 \citep{Salatino2012} for a modified blackbody spectral energy
distribution.

\begin{figure}
\centering
\includegraphics[width=0.86\linewidth]{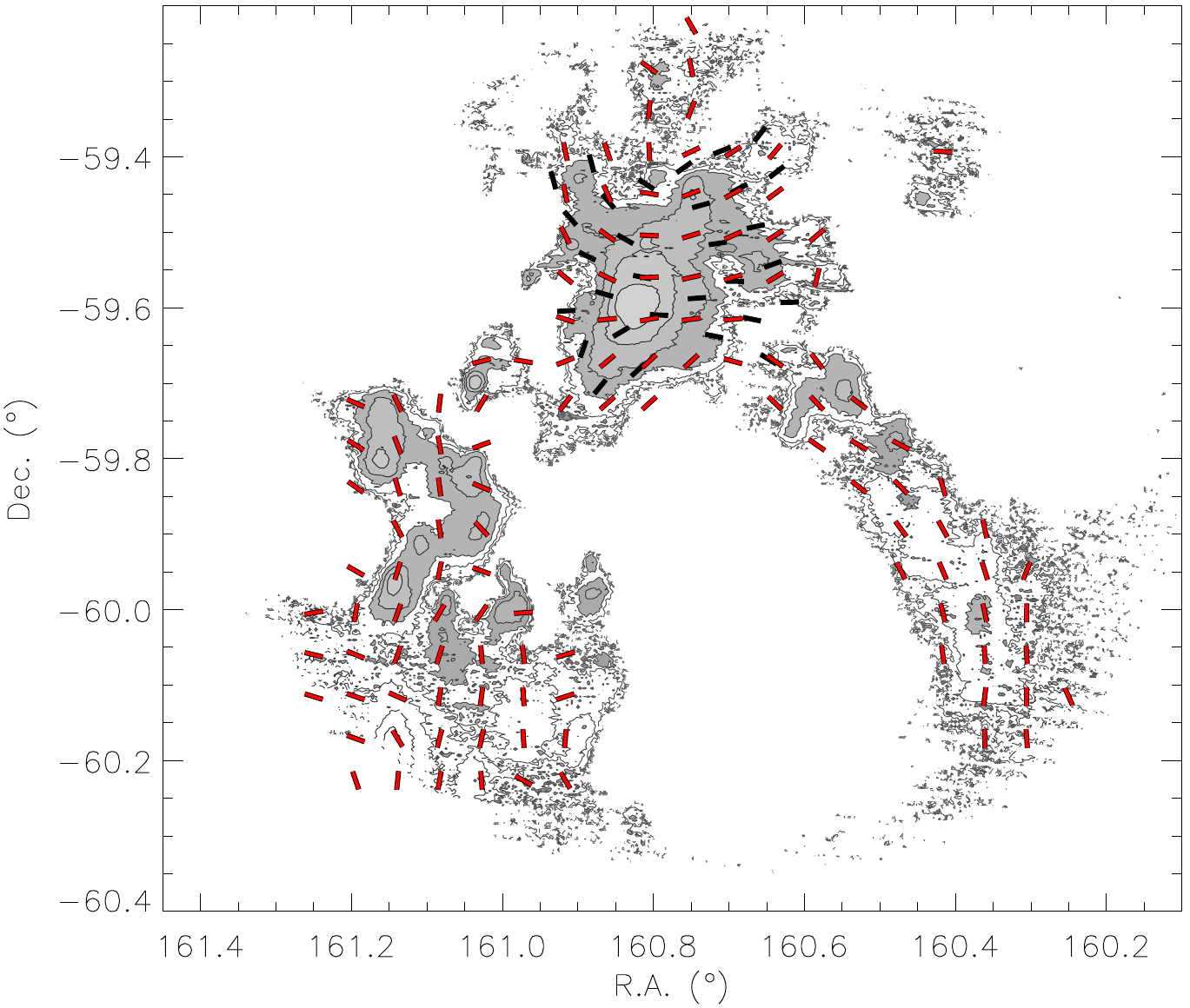}
\caption{Preliminary BLASTPol map at 500\,$\mu$m of the Carina
Nebula, a giant molecular cloud approximately centred at
coordinates $[10^{\rm h} 42^{\rm m} 35^{\rm s}, -59^{\circ}
42\arcmin 15\arcsec]$. The intensity contour levels are shown in
the background. The pseudo-vectors indicate the inferred magnetic
field direction: the BLASTPol measurements (red) are in excellent
agreement with those from the SPARO polarimeter at 450\,$\mu$m
\citep[black;][]{Li2006}. This map should not be considered of any
scientific value as no absolute calibration has been applied to
the intensity contours; the map is only shown as a proof of
concept for the map-maker, which includes all the corrections due
to HWP non-idealities discussed in this work. Here we adopt
$\overline{\beta}^{\rm 500\,\mu m}_{\rm ea} = 1.7$, calculated
assuming a spectral index for the dust emissivity of 1.37
\citep{Salatino2012}.} \label{fig:CarinaNeb_500um_pol_map}
\end{figure}

The map in Fig.~\ref{fig:CarinaNeb_500um_pol_map} is  presented as
contour levels of the intensity map $I$, on which we superimpose
pseudo-vectors indicating the inferred magnetic field direction on
the sky \citep[assumed to be perpendicular to the measured
polarisation direction; see][]{Lazarian2007}. The sky polarisation
angle is given by $\phi=\frac{1}{2}\arctan\frac{U}{Q}$. Because
the absolute flux calibration has not been finalised yet, we
choose not to report here the intensity values corresponding to
each contour level. This map should not be considered of any
scientific value as it is not calibrated in flux. Nonetheless, we
find a remarkable agreement between the BLASTPol polarisation
directions at 500\,$\mu$m and those produced by the Submillimeter
Polarimeter for Antarctic Remote Observations
\citep[SPARO;][]{Novak2003} at 450\,$\mu$m, which are shown in
Fig.~1 of \citet{Li2006} and whose pseudo-vectors are reported in
Fig.~\ref{fig:CarinaNeb_500um_pol_map} for comparison. Here the
original BLASTPol $\left[I,Q,U\right]$ maps have been smoothed
with a kernel of 4\arcmin\ (FWHM) for a more direct comparison
with the SPARO data.

\section{Concluding Remarks}\label{sec:concl_HWP}

The goal of the first part of this work was to identify and
measure the parameters that fully characterise the spectral
performance of the linear polarisation modulator integrated in the
BLASTPol instrument, a cryogenic achromatic HWP.

We have described in detail the design and manufacturing process
of a five-slab sapphire HWP, which is, to our knowledge, the most achromatic built to date at (sub)mm wavelengths. In the same
context, we have provided a useful collection of spectral data
from the literature for the sapphire absorption coefficient at
cryogenic temperatures.

Using a polarising FTS, we have fully characterised the spectral
response of the anti-reflection coated BLASTPol HWP at room
temperature and at 120\,K; we have acquired data
cubes by measuring spectra while rotating the HWP to produce the
polarisation modulation.

The cold dataset contains measurements in both co-pol and
cross-pol configurations; we have used these two data cubes to
estimate 9 out of 16 elements of the HWP Mueller matrix as a
function of frequency. We have developed an ad-hoc Monte Carlo
algorithm that returns for every frequency the best estimate of
each matrix element and the associated error, which is a
combination of the uncertainty on the measured spectra and a
random jitter on the rotation angle.

We have measured how the position of the equivalent axes of the
HWP, $\beta_{\rm ea}$, changes as a function of frequency, an
effect that is inherent to any achromatic design. Once this
dependence is accounted for in the Monte Carlo, and a correction
is implemented for the residual absorption from sapphire, the
Mueller matrix of the HWP approaches that of an ideal HWP, at all
wavelengths of interest. In particular, the (band-averaged)
off-diagonal elements are always consistent with zero within
2\,$\sigma$ and the modulus of the three diagonal coefficients is
always $>$\,0.8. Therefore, we have introduced in the BLASTPol
map-making algorithm the band-integrated values of $\beta_{\rm
ea}$ as an additional parameter in the evaluation of the
polarisation angle. To first order, this approach allows us to
account for most of the non-idealities in the HWP.

We have investigated the impact of input sources with different
spectral signatures on $\beta_{\rm ea}$ and on the HWP Mueller
matrix coefficients. We find that the HWP transmission and
modulation efficiency are very weakly dependent on the spectral
index of the input source, whereas the position of the equivalent
axes of the sapphire plate stack is more significantly affected.
This latter dependence, if neglected, may lead to an arbitrary
rotation of the retrieved polarisation angle on the sky of
magnitude $2\,\overline{\beta}_{\rm ea} =10$--15$^{\circ}$
(3--5$^{\circ}$) at 250 (500)\,$\mu$m. The 350\,$\mu$m band,
however, is minimally perturbed by this effect.

In principle, the measured Mueller matrix can be used to generate
a synthetic time-ordered template of the polarisation modulation
produced by the HWP as if it were continuously rotated at a
mechanical frequency $f=\omega\,t$. Continuous rotation of the HWP
allows the rejection of all the noise components modulated at
harmonics different than 4\,$f$ (synchronous demodulation) and is
typically employed by experiments optimised to measure the
polarisation of the CMB
\citep[e.g.,][]{Johnson2007,Reichborn2010}. In such experiments,
the HWP modulation curve leaves a definite synchronous imprint on
the time-ordered bolometer data streams, hence it is
of utter importance to characterise the template and remove it
from the raw data. However, a time-ordered HWP template would be
of no use to a step-and-integrate experiment such as BLASTPol,
whose timelines are not dominated by the HWP synchronous signal.

We have measured the phase shift of the HWP across the wavelength
range of interest to be within 5$^{\circ}$ of the ideal 180$^{\circ}$ for the central BLASTPol band, and within 15$^{\circ}$ for the side bands. This is due to a combination of alignment errors of the sapphire substrates, which are hard to avoid in the manufacture of a five-slab stack, and their lower than ideal thickness. However, the modulation efficiency of the
HWP is only mildly affected by this departure from ideality, being
above 98\% in all three BLASTPol bands. Moreover, departures of
similar amplitude are not uncommon for HWPs at (sub)mm
wavelengths.

The goal of the second part of this work was to include the
measured non-idealities of the HWP as-built in a map-making
algorithm. We have focused on the implementation of a naive
binning technique for the case of BLASTPol, under the assumption
of white and uncorrelated noise. As a proof of concept, we have
presented a preliminary polarisation map for one of the scientific
targets observed by BLASTPol during its first Antarctic flight,
completed in January 2011. The inferred direction for the local
magnetic field in the Carina Nebula star-forming region is in
excellent agreement with the results obtained by \citet{Li2006}
with the SPARO instrument. The empirical approach presented in
this paper will help improve the accuracy on
astronomical measurements of the polarisation angle on the sky at
submm wavelengths.

\section*{Acknowledgments}

The BLASTPol collaboration acknowledges the support of NASA
through grant numbers NAG5-12785, NAG5-13301 and NNGO-6GI11G, the
Canadian Space Agency (CSA), the Science and Technology Facilities
Council (STFC), Canada's Natural Sciences and Engineering Research
Council (NSERC), the Canada Foundation for Innovation, the Ontario
Innovation Trust, the Puerto Rico Space Grant Consortium, the
Fondo Istitucional para la Investigacion of the University of
Puerto Rico, the National Science Foundation Office of Polar
Programs, and the Canadian Institute for Advanced Research.

\bibliographystyle{mn2e}
\bibliography{refs}

\label{lastpage}
\end{document}